\documentclass[11pt]{article}

\newsavebox{\foobox}
\newcommand{\slantbox}[2][0]{\mbox{%
        \sbox{\foobox}{#2}%
        \hskip\wd\foobox
        \pdfsave
        \pdfsetmatrix{1 0 #1 1}%
        \llap{\usebox{\foobox}}%
        \pdfrestore
}}
\newcommand\unslant[2][-.25]{\slantbox[#1]{$#2$}}

\newcommand{\mmu}{\text{\unslant\mu}}
\newcommand{\mpi}{\text{\unslant[-.18]\pi}}
\newcommand{\mdelta}{\text{\unslant[-.18]\delta}}
\newcommand{\mzeta}{\text{\unslant[-.15]\zeta}}

\renewcommand\textfraction{.05}

\usepackage[left=2cm, right=2cm, top=2.5cm, bottom=2.5cm]{geometry}
\usepackage[x11names]{xcolor}
\usepackage{fancyhdr, amssymb, cancel, amsmath, graphicx, pgfplots, tikz}
\usepackage{isomath}

\usetikzlibrary{shadows}

\newcommand{\stylecolor}{blue!50!black}

\usepackage[labelfont={bf,sf, color=\stylecolor}, margin={1.5cm,0cm}]{caption}

\usepackage[colorlinks=true, urlcolor=\stylecolor!70!white, linkcolor=\stylecolor, citecolor=\stylecolor!70!white, hyperindex=true, linktocpage=true]{hyperref}

\usepackage[explicit]{titlesec}

\newcommand*\sectionlabel{}
\titleformat{\section}
  {\gdef\sectionlabel{}
   \Large\bfseries\scshape}
  {\gdef\sectionlabel{\thesection }}{0pt}
  {\begin{tikzpicture}[remember picture,overlay]
       \end{tikzpicture}
  }
\titlespacing*{\section}{0pt}{0pt}{0pt}

\newcommand*\subsectionlabel{}
\titleformat{\subsection}
  {\gdef\subsectionlabel{}
   \large\bfseries\scshape}
  {\gdef\subsectionlabel{\thesubsection  }}{0pt}
  {\begin{tikzpicture}[remember picture,overlay]
    	\draw (-0.15, 0.02) node[right] {\color{\stylecolor} \textsf{\subsectionlabel}};
	\draw (1.25, 0) node[right] {\color{\stylecolor} \textsf{#1}};
	\fill[color=\stylecolor] (1,-0.25) rectangle (1.1, 0.25);
       \end{tikzpicture}
  }
\titlespacing*{\subsection}{0pt}{10pt}{10pt}

\newcommand*\subsubsectionlabel{}
\titleformat{\subsubsection}
  {\gdef\subsubsectionlabel{}
   \bfseries\scshape}
  {\gdef\subsubsectionlabel{\thesubsubsection.\ \  }}{0pt}
  {\begin{tikzpicture}[remember picture,overlay]
    	\draw (-0.15, 0) node[right] {\color{\stylecolor} \textsf{\subsubsectionlabel#1}};
       \end{tikzpicture}
  }
\titlespacing*{\subsubsection}{0pt}{7pt}{7pt}

\pgfplotsset{every axis legend/.append style={at={(1.02,1)},anchor=north west}}

\begin{document}

\allowdisplaybreaks

\setcounter{totalnumber}{5}
\renewcommand\textfraction{.1}

\pagestyle{fancy}
\renewcommand{\headrulewidth}{0pt}
\fancyhead{}

\fancyfoot{}
\fancyfoot[C] {\textsf{\textbf{\thepage}}}

\begin{equation*}
\begin{tikzpicture}
\draw (\textwidth, 0) node[text width = \textwidth, right] {\color{white} easter egg};
\end{tikzpicture}
\end{equation*}

\begin{equation*}
\begin{tikzpicture}
\draw (0.5\textwidth, -3) node[text width = \textwidth] {\huge  \textsf{\textbf{Transport in   inhomogeneous quantum critical fluids \vspace{0.07in} \\   and in the Dirac fluid in graphene }} };
\end{tikzpicture}
\end{equation*}
\begin{equation*}
\begin{tikzpicture}
\draw (0.5\textwidth, 0.1) node[text width=\textwidth] {\large \color{black} \textsf{Andrew Lucas,}$^{\color{\stylecolor} \mathsf{a}}$ \textsf{Jesse Crossno,}$^{\color{\stylecolor} \mathsf{a,b}}$ \textsf{Kin Chung Fong,}$^{\color{\stylecolor} \mathsf{c}}$ \textsf{Philip Kim,}$^{\color{\stylecolor} \mathsf{a,b}}$ \textsf{and Subir Sachdev}$^{\color{\stylecolor} \mathsf{a,d}}$};
\draw (0.5\textwidth, -0.5) node[text width=\textwidth] { $^{\color{\stylecolor} \mathsf{a}}$  \small\textsf{Department of Physics, Harvard University, Cambridge, MA 02138, USA}};
\draw (0.5\textwidth, -1) node[text width=\textwidth] { $^{\color{\stylecolor} \mathsf{b}}$  \small\textsf{John A. Paulson School of Engineering and Applied Sciences, Harvard University, Cambridge, MA 02138, USA}};
\draw (0.5\textwidth, -1.5) node[text width=\textwidth] { $^{\color{\stylecolor} \mathsf{c}}$  \small\textsf{Quantum Information Processing Group, Raytheon BBN Technologies,  Cambridge,
MA 02138, USA}};
\draw (0.5\textwidth, -2) node[text width=\textwidth] { $^{\color{\stylecolor} \mathsf{d}}$  \small\textsf{Perimeter Institute for Theoretical Physics, Waterloo, Ontario N2L 2Y5, Canada}};
\end{tikzpicture}
\end{equation*}
\begin{equation*}
\begin{tikzpicture}
\draw (0, -13.1) node[right, text width=0.5\paperwidth] {\textsf{Corresponding author: }\texttt{lucas@fas.harvard.edu}};
\draw (\textwidth, -13.1) node[left] {\textsf{\today}};
\end{tikzpicture}
\end{equation*}
\begin{equation*}
\begin{tikzpicture}
\draw[very thick, color=\stylecolor] (0.0\textwidth, -5.75) -- (0.99\textwidth, -5.75);
\draw (0.12\textwidth, -6.25) node[left] {\color{\stylecolor}  \textsf{\textbf{Abstract:}}};
\draw (0.53\textwidth, -6) node[below, text width=0.8\textwidth, text justified] {\small We develop a general hydrodynamic framework for computing direct current  thermal and electric transport in a  strongly interacting finite temperature quantum system near a Lorentz-invariant quantum critical point.  
Our framework is non-perturbative in the strength of long wavelength fluctuations in the background charge density of the electronic fluid,  and requires the rate of electron-electron scattering
to be faster than the rate of electron-impurity scattering.
We use this formalism to compute transport coefficients in the Dirac fluid in clean samples of graphene near the charge neutrality point, and find results insensitive to long range Coulomb interactions.  Numerical results are compared to recent experimental data on thermal and electrical conductivity in the Dirac fluid in graphene and substantially improved quantitative agreement over existing hydrodynamic theories is found.  We comment on the interplay between the Dirac fluid and acoustic and optical phonons, and qualitatively explain experimentally observed effects.  Our work paves the way for quantitative contact between experimentally realized condensed matter systems and the wide body of high energy inspired theories on transport in  interacting many-body quantum systems.};
\end{tikzpicture}
\end{equation*}

\tableofcontents

\titleformat{\section}
  {\gdef\sectionlabel{}
   \Large\bfseries\scshape}
  {\gdef\sectionlabel{\thesection }}{0pt}
  {\begin{tikzpicture}[remember picture,overlay]
	\draw (1, 0) node[right] {\color{\stylecolor} \textsf{#1}};
	\fill[color=\stylecolor] (0,-0.35) rectangle (0.7, 0.35);
	\draw (0.35, 0) node {\color{white} \textsf{\sectionlabel}};
       \end{tikzpicture}
  }
\titlespacing*{\section}{0pt}{15pt}{15pt}

\begin{equation*}
\begin{tikzpicture}
\draw[very thick, color=\stylecolor] (0.0\textwidth, -5.75) -- (0.99\textwidth, -5.75);
\end{tikzpicture}
\end{equation*}

\section{Introduction}
Over a half century ago, the theory of electronic transport in ``standard" metals such as iron and copper was developed.  The key pillar of this approach is the validity of Fermi liquid theory, which states that the interacting electrons in solids form nearly free-streaming quasiparticles \cite{pines}.  At finite temperature, these quasiparticles form a weakly interacting quantum gas which is well described by quantum kinetic theory.   The transport properties of these quantum gases are by now very well understood.   A particularly important property of Fermi liquids is the Wiedemann-Franz law, which states that\footnote{Below we have assumed that the charge of the quasiparticles is $\pm e$, with $e$ the charge of the electron -- this is essentially always the case.} \begin{equation}
\mathcal{L} \equiv \frac{\kappa}{\sigma T} = \frac{\mpi^2}{3} \frac{k_{\mathrm{B}}^2}{e^2} \equiv \mathcal{L}_{\mathrm{WF}}.
\end{equation}
Here $\kappa$ is the electronic contribution to thermal conductivity, $\sigma$ is the electrical conductivity, $T$ is the temperature, and $\mathcal{L}$ is the Lorenz ratio.   Implicit in the above equation is that the dominant interactions are between impurities or phonons and quasiparticles, and in most metals this is true:  the interaction time between quasiparticles is typically $10^4$ times longer than the interaction times between quasiparticles and impurities or phonons \cite{ashcroft}.   

Also over a half century ago,  a study of the consequences of hydrodynamic behavior on correlation functions and transport in interacting quantum systems was initiated \cite{kadanoff}.    Hydrodynamics is a framework for understanding the collective motion of the quasiparticles in a solid, or any other interacting quantum or classical system, so long as the microscopic degrees of freedom reach thermal equilibrium locally.   In a solid, this interaction time must be the fastest time scale in the problem to see hydrodynamic behavior, but since quasiparticles in a Fermi liquid interact with each other only very weakly, observing hydrodynamics in electron fluids is notoriously hard.  Even in the purest metals where hydrodynamic behavior can be observed, such as in GaAs \cite{molenkamp, weber, lilly},  the resulting fluid is often a Fermi liquid.  The resulting dynamics is the fluid dynamics of (quantum) gases.  More recent theoretical work on hydrodynamics in Fermi liquids 
includes \cite{andreev, succiturb, tomadin, vignale, polini, levitovhydro}, and recent experimental work includes \cite{bandurin, mackenzie}.

Fermi liquid theory is known to fail in a variety of experimentally realized metals in two or more spatial dimensions -- most famous among these is the strange metal phase of the cuprate superconductors \cite{vandermarel, hussey, SSBK11}  which does not have quasiparticle excitations.  A slightly more theoretically controlled  and better understood example  of a state of quantum matter without quasiparticles is the quasi-relativistic Dirac fluid in the semimetal graphene.   The Dirac fluid, which effectively lives in two spatial dimensions, has also been argued to be strongly interacting at experimentally achievable temperatures \cite{schmalian, muller1, muller4, andrei} due to ineffective Coulomb screening \cite{lanzara}.    Although it is separated from the Fermi liquid by a crossover, and not a (thermal) phase transition, its proximity to a (simple) zero temperature quantum critical point at charge neutrality means that the phenomenology of the Dirac fluid is expected to differ strongly from Fermi liquid theory.   Due to the high spatial dimensionality,\footnote{Quantum dynamics in one dimension,  which is often integrable, is described using very different techniques and has qualitatively distinct features.} the development of a predictive quantitative theory of these systems is notoriously hard.   A major theme in recent work has been quantum criticality \cite{damle97,sachdev}, which opens up the possibility for borrowing powerful techniques from high energy physics,  but even in this case very little is known about experimentally relevant regime of finite temperature and density.    One of the only remaining techniques for understanding these systems is hydrodynamics, as many features of hydrodynamics are universal and model independent, and the strongly interacting quantum physics is captured entirely by the coefficients in otherwise classical differential equations.  Such fluids are quantum analogues of classical liquids such as water, which are strongly interacting (albeit with negligible quantum entanglement) insofar as they do not admit a controllable description via kinetic theory.   Furthermore, it has been shown \cite{hkms} that strongly interacting quantum critical fluids have a somewhat different hydrodynamic description than the canonical Fermi liquids described above, and this can lead to very different hydrodynamic properties,  including in transport \cite{hkms, muller1, muller4, muller2, muller3, foster, dsz}, as we will review in this paper.

Using novel techniques to measure thermal transport \cite{fong, fong2, crossno2},  the Dirac fluid has finally been  observed in monolayer graphene, and evidence for its hydrodynamic behavior  has emerged \cite{crossno}, as we will detail.  However, existing theories of hydrodynamic transport are not consistent with the simultaneous density dependence in experimentally measured thermal and electrical conductivities.   In this paper, we improve upon the hydrodynamic theory of \cite{hkms}, describe carefully effects of finite density, and develop a non-perturbative relativistic hydrodynamic theory of transport in electron fluids near a quantum critical point.   Under certain assumptions about the equations of state of the Dirac fluid, our theory is quantitatively consistent with experimental observations.   The techniques we employ are included in the framework of \cite{lucas}, which developed a hydrodynamic description of transport in relativistic fluids with long wavelength disorder in the chemical potential.   \cite{lucas} was itself inspired by recent progress employing the AdS/CFT correspondence to understand quantum critical transport in strange metals \cite{btv, dsz, herzog, lss, lucas1501, davison15, blake2, donos1506,  grozdanov}, but as we will discuss, this theory is also well suited to describe the physics of graphene.

\subsection{Summary of Results}
\begin{figure}[t]
\centering
\includegraphics[width=7in]{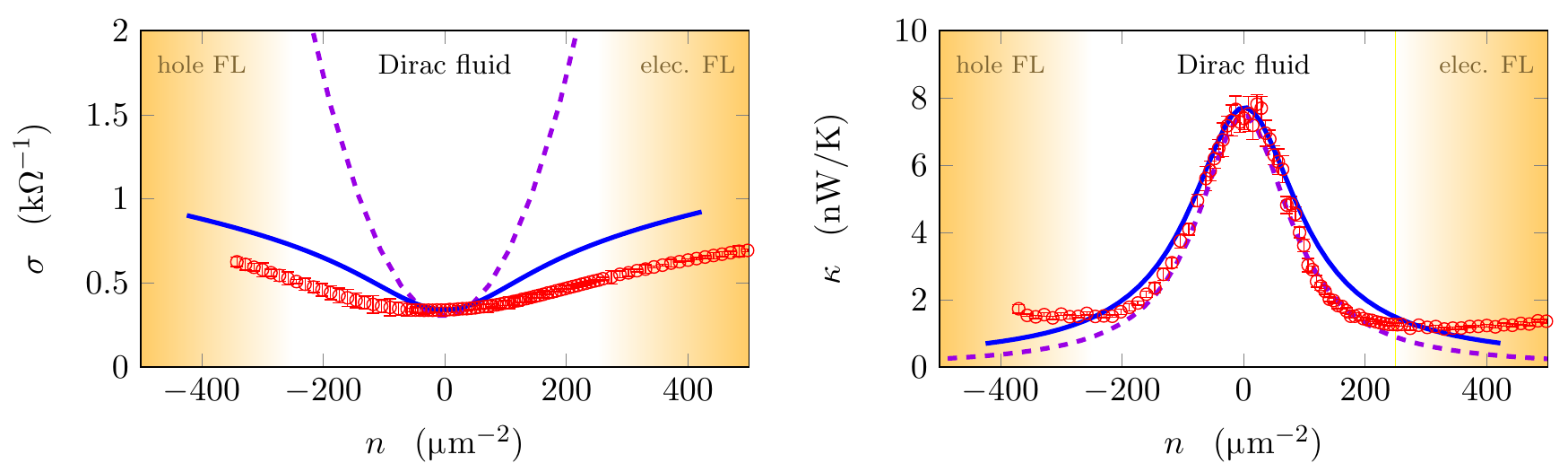}
\caption{A comparison of our hydrodynamic theory of transport with the experimental results of \cite{crossno} in clean samples of graphene at $T=75$ K.   We study the electrical and thermal conductances at various charge densities $n$ near the charge neutrality point.  Experimental data is shown as circular red data markers, and numerical results of our theory, averaged over 30 disorder realizations, are shown as the solid blue line.   Our theory assumes the equations of state described in (\ref{numericmain}) with the parameters $C_0\approx 11$, $C_2\approx 9$,  $C_4\approx 200$,  $\eta_0\approx 110$, $\sigma_0\approx 1.7$,  and (\ref{numericmain2}) with $u_0 \approx 0.13$.   The yellow shaded region shows where Fermi liquid behavior is observed and the Wiedemann-Franz law is restored, and our hydrodynamic theory is not valid in or near this regime.   We also show the predictions of (\ref{hkmseq}) as dashed purple lines, and have chosen the 3 parameter fit to be optimized for $\kappa(n)$. }
\label{mainfig}
\end{figure}

The recent experiment \cite{crossno} reported order-of-magnitude violations of the Wiedemann-Franz law.   The results were compared with the standard theory of hydrodynamic transport in quantum critical systems \cite{hkms}, which predicts that \begin{subequations}\label{hkmseq}\begin{align}
\sigma(n) &= \sigma_{\textsc{q}} + \frac{e^2v_{\mathrm{F}}^2n^2\tau}{\mathcal{H}}, \\
\kappa(n) &= \frac{v_{\mathrm{F}}^2\mathcal{H}\tau}{T}  \frac{\sigma_{\textsc{q}}}{\sigma(n)},
\end{align}\end{subequations}
where $e$ is the electron charge, $s$ is the entropy density, $n$ is the charge density (in units of $\mathrm{length}^{-2}$),  $\mathcal{H}$ is the enthalpy
 density,  $\tau$ is a momentum relaxation time, and $\sigma_{\textsc{q}}$ is a quantum critical effect,  whose existence is a new effect in the hydrodynamic gradient expansion of a relativistic fluid.   Note that up to $\sigma_{\textsc{q}}$,  $\sigma(n)$ is simply described by Drude physics.  The Lorenz ratio then takes the general form \begin{equation}
\mathcal{L}(n) = \frac{\mathcal{L}_{\mathrm{DF}}}{(1+(n/n_0)^2)^2},  \label{hkmsLeq}
\end{equation}
 where
\begin{subequations}\begin{align}
\mathcal{L}_{\mathrm{DF}} &= \frac{v_{\mathrm{F}}^2\mathcal{H}\tau}{T^2\sigma_{\textsc{q}}}, \\
n_0^2 &= \frac{\mathcal{H} \sigma_{\textsc{q}}}{e^2 v_{\mathrm{F}}^2 \tau}.
\end{align}\end{subequations}
$\mathcal{L}(n)$ can be parametrically larger than $\mathcal{L}_{\mathrm{WF}}$ (as $\tau\rightarrow\infty$ and $n\ll n_0$),  and much smaller ($n\gg n_0$).    Both of these predictions were observed in the recent experiment,  and fits of the measured $\mathcal{L}$ to (\ref{hkmsLeq}) were quantitatively consistent,  until large enough $n$ where Fermi liquid behavior was restored.    However, the experiment also found that the conductivity did not grow rapidly away from $n=0$ as predicted in (\ref{hkmseq}), despite a large peak in $\kappa(n)$ near $n=0$,  as we show in Figure \ref{mainfig}.   Furthermore, the theory of \cite{hkms} does not make clear predictions for the temperature dependence of $\tau$, which determines $\kappa(T)$.

\begin{figure}[t]
\centering
\includegraphics[width=7in]{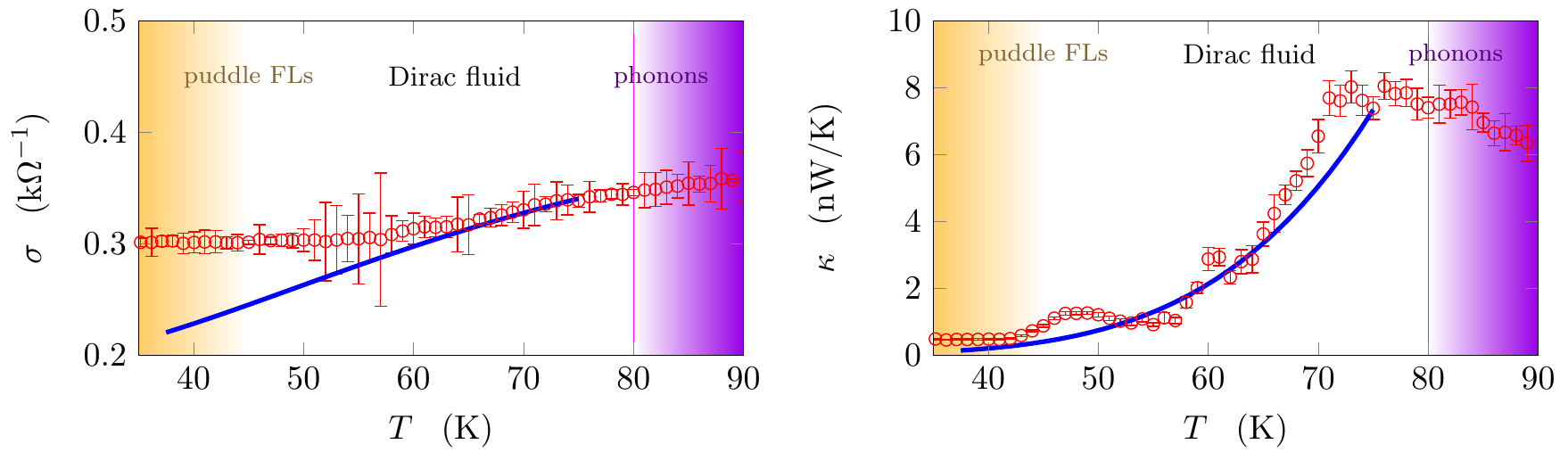}
\caption{A comparison of our hydrodynamic theory of transport with the experimental results of \cite{crossno} in clean samples of graphene at the charge neutrality point ($n=0$).  We use no new fit parameters compared to Figure \ref{mainfig}.   The yellow shaded region denotes where Fermi liquid behavior is observed; the purple shaded region denotes the likely onset of electron-phonon coupling.}
\label{mainfigT}
\end{figure}

In this paper, we argue that there are two related reasons for the breakdown of (\ref{hkmseq}).  One is that the dominant source of disorder in graphene -- fluctuations in the local charge density, commonly referred to as charge puddles \cite{yacoby2007, sarmachargepuddles, crommie, xue} --  are not perturbatively weak, and therefore a non-perturbative treatment of their effects is necessary.\footnote{See \cite{sarma1, sarma2} for a theory of electrical conductivity in charge puddle dominated graphene at low temperatures.}   The second is that the parameter $\tau$, even when it is sharply defined, is intimately related to both the viscosity and to $n$, and this $n$ dependence is neglected when performing the  fit to (\ref{hkmseq}) in Figure \ref{mainfig}.    We develop a non-perturbative hydrodynamic theory of transport which relies on neither of the above assumptions, and gives us an explicit formula for $\tau$ in the limit of weak disorder.   The key assumption for the validity of our theory is that the size of the charge puddles is comparable to or larger than the electron interaction length scale, which is about 100 nm.   Experimental evidence suggests this is  marginally true in graphene samples mounted on hexagonal boron nitride \cite{xue}, as was done in \cite{crossno}.   Although we cannot analytically solve our theory non-perturbatively, we perform numerical computations of the transport coefficients in disordered fluids, and compare the results to the experimental data in Figure \ref{mainfig}.  Our simultaneous fit to $\kappa(n)$ and $\sigma(n)$ shows improved quantitative agreement with both sets of data in the Dirac fluid regime.    We further compare in Figure \ref{mainfigT} the temperature dependence of $\kappa$ and $\sigma$ between our numerics and the experiment, using no new fitting parameters, and find satisfactory quantitative agreement in the Dirac fluid regime.  

\begin{figure}[t]
\centering
\includegraphics[width=3.3in]{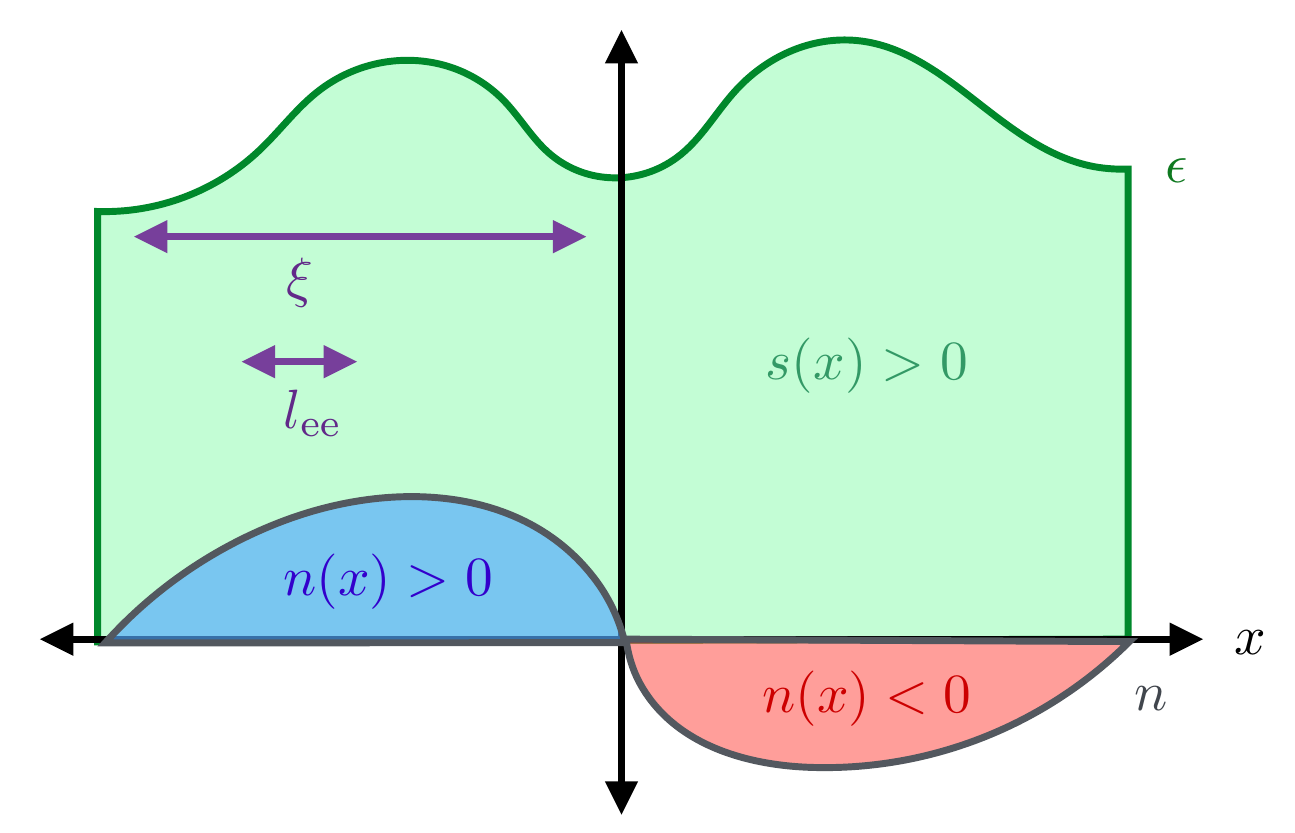}
\caption{A cartoon of a nearly quantum critical fluid where our hydrodynamic description of transport is sensible.   The local chemical potential $\mu(\mathbf{x})$ always obeys $|\mu| \ll k_{\mathrm{B}}T$,  and so the entropy density $s/k_{\mathrm{B}}$ is much larger than the charge density $|n|$;  both electrons and holes are everywhere excited, and the energy density $\epsilon$ does not fluctuate as much relative to the mean.  Near charge neutrality the local charge density flips sign repeatedly.   The correlation length of disorder $\xi$ is much larger than $l_{\mathrm{ee}}$, the electron-electron interaction length.}
\label{hydrofig}
\end{figure}

Figure \ref{hydrofig} shows a cartoon of the regime of validity of our hydrodynamic theory.   The fact that the charge puddles may be substantial, while the entropy and energy densities are much more constant,  helps to explain why the perturbative description of transport is much better for $\kappa$ than $\sigma$,  as the perturbative approach works well in a nearly homogeneous fluid.  In coming years the quality of graphene samples will improve, and the charge puddle size may grow larger than 100 nm, allowing us to observe the clean hydrodynamic limit described by (\ref{hkmseq}).   As present day samples are just clean enough to observe hydrodynamics, our determination of the equations of state should be understood as preliminary.


Although the focus of this paper is on the Dirac fluid in graphene, this is because of the experimental motivation for this work.   Our theory has broader validity, and we will introduce it in the more general context of transport in a disordered electronic fluid near a quantum critical point with manifest Lorentz invariance, with the microscopic Fermi velocity $v_{\mathrm{F}}$ playing the role of the speed of light.  The Dirac fluid is not strictly Lorentz invariant, but we will justify the validity of our approach even in this system.   While the Dirac fluid in graphene is currently the only experimentally realized strongly interacting condensed matter system with evidence for electronic hydrodynamics \cite{crossno},  in the future surface states in topological insulators in three spatial dimensions  may host strongly interacting electron fluids \cite{galitski}.  Strongly interacting three dimensional materials including Weyl semimetals \cite{soljacic, syxu, bqlv} may also give rise to novel phenomena relevant for transport \cite{nielsen, spivakson}.

\subsection{Outline}
The outline of this paper is as follows.  We briefly review the definitions of transport coefficients in Section \ref{sectransco}.   In Section \ref{sechydro} we develop a theory of hydrodynamic transport in the electron fluid, assuming that it is Lorentz invariant.  We discuss the peculiar case of the Dirac fluid in graphene in Section \ref{secgraphene}, and argue that deviations from Lorentz invariance are small.   We describe the results of our numerical simulations of this theory in Section \ref{secnum}, and directly compare our simulations with recent experimental data from graphene \cite{crossno}.  The experimentally relevant effects of phonons are qualitatively described in Section \ref{secphonon}.  We conclude the paper with a discussion of future experimental directions.   Appendices contain technical details of our theory.

In this paper we use index notation for vectors and tensors.  Latin indices $ij\cdots$ run over spatial coordinates $x$ and $y$;  Greek indices $\mu\nu\cdots$ run over time $t$ as well.   We will denote the time-like coordinate of $A^\mu$ as $A^t$.  Indices are raised and lowered with the Minkowski metric $\eta^{\mu\nu} \equiv \mathrm{diag}(-1,1,1)$.  The Einstein summation convention is always employed.

\section{Transport Coefficients}\label{sectransco}
Let us begin by defining the thermoelectric response coefficients of interest in this paper.   Suppose that we drive our fluid by a spatially uniform, externally applied, electric field $ E_i$ (formally, an electrochemical potential gradient), and a temperature gradient $-\partial_i T$.  We will refer to $-\partial_j T$ as $T\zeta_j$, with $\zeta_j = -T^{-1}\partial_j T$, for technical reasons later.   As is standard in linear response theory, we decompose these perturbations into various frequencies, and focus on the response at a single frequency $\omega$.   Time translation invariance implies that the (uniformly) spatially averaged charge current $\langle  J_i \rangle$ and the spatially averaged heat current $\langle  Q_i \rangle$ are also periodic in time of frequency $\omega$, and are related to $ E_i$ and $ \zeta_i$ by the thermoelectric transport coefficients: \begin{equation}
\left( \begin{array}{c} \langle  J_i \rangle  \\ \langle  Q_i\rangle \end{array}\right) \mathrm{e}^{-\mathrm{i}\omega t} =  \left( \begin{array}{cc} \sigma_{ij}(\omega)   &\  T\alpha_{ij}(\omega) \\  T\bar\alpha_{ij}(\omega)   &\ T\bar\kappa_{ij}(\omega)  \end{array}\right)\left( \begin{array}{c}  E_j \\  \zeta_j  \end{array}\right) \mathrm{e}^{-\mathrm{i}\omega t}.   \label{transeq}
\end{equation}
 In a typical disordered system, we expect that $\sigma_{ij}$, $\alpha_{ij}$, $\bar\alpha_{ij}$ and $\bar\kappa_{ij}$ are all proportional to $\mdelta_{ij}$.   In our numerics, finite size effects introduce some anisotropy; our theory is valid in this more general scenario.

In fact, (\ref{transeq}) is somewhat subtle.   Implicit in the definitions of the transport coefficients are a set of boundary conditions.   In the definitions in (\ref{transeq}), we have assumed that we tune $E_i$ and $ \zeta_i$, and then measure $ J_i$ and $Q_i$.   However, usually in experiments one fixes $J_i$, as electronic measurements are far easier to control.  One then can fix either $E_i$ or $\zeta_i$. So while it is straightforward to measure $\sigma_{ij}$ by setting $\zeta_i=0$,  one measures not $\bar{\kappa}_{ij}$ but instead $\kappa_{ij}$, defined as \begin{equation}
\left.\langle  Q_i\rangle\right|_{\langle  J_i\rangle = 0} = T\kappa_{ij} \zeta_j.
\end{equation}Straightforward manipulations give that $\sigma_{ij}E_j = -T\alpha_{ij}\zeta_j$, and therefore that\begin{equation}
\kappa_{ij} = \bar\kappa_{ij} - T\bar\alpha_{ik}\sigma^{-1}_{kl}\alpha_{lj}.
\end{equation}
These definitions are general and independent of our hydrodynamic theory.

\section{Relativistic Hydrodynamics}\label{sechydro}
We now develop a theory of relativistic hydrodynamics of the electronic plasma in a disordered metal, where the disorder is introduced by a spatially dependent chemical potential $\mu_0(\mathbf{x})$.    So long as the length scale $\xi \sim |\mu_0|/ |\partial_x \mu_0|$ over which this function varies is much larger than the electron-electron scattering length $l_{\mathrm{ee}} \sim \hbar v_{\mathrm{F}} /k_{\mathrm{B}}T$, it is sensible to treat the fluid as locally homogeneous, with parameters such as energy density and viscosity locally being functions of $\mu_0$ alone.  This external chemical potential acts as an external source of energy and momentum for the electronic plasma, and can be sourced by lattice defects or impurities, either in the (semi)metal itself, or in the substrate it is placed on, for two-dimensional materials such as graphene \cite{yacoby2007, xue}.   Our theory here is  analogous to \cite{lucas}, and similar to the earlier work \cite{andreev} in non-relativistic fluids.   However, \cite{lucas} focused mostly on the mathematical consequences of relativistic hydrodynamics, particularly in regards to holographic models.   Our focus here is on practical consequences in realistic quantum critical fluids where $\mu \ll k_{\mathrm{B}}T$, and where the equations of state are tightly constrained by scale invariance (see Appendix \ref{appthermo}).

Previous theories of hydrodynamic transport assumed that disorder was parametrically weak, and so momentum is a nearly conserved quantity \cite{hkms, muller1}.    Such theories can be shown to be a perturbative limit of the more general approach that we advocate below: see \cite{lucas} and Appendix \ref{appmom}.    However, near the charge-neutrality point, non-perturbative effects can become important \cite{lucas}.  Since this is the regime where \cite{crossno} observed evidence for hydrodynamic behavior, it is necessary to treat transport in the charge-neutral fluid carefully and to study non-perturbative physics.   We begin with a general discussion of the equations of state of a relativistic plasma, and then outline our non-perturbative hydrodynamic description of  transport.

Though our focus in this paper is on the case of two spatial dimensions, it is straightforward to generalize our theory to higher dimensions.

\subsection{Hydrodynamic Equations}
Let us review the structure of relativistic hydrodynamics, which was derived carefully in \cite{hkms}.  Hydrodynamics is a general framework which describes the relaxation of an ergodic and locally thermalizing (classical or quantum) system  to global thermal equilibrium, or as close to global equilibrium as boundary conditions or external sourcing allow.
The assumption of local thermalization implies that the only quantities with dynamics on long time scales (compared to the local thermalization time $l_{\mathrm{ee}}/v_{\mathrm{F}}$)  are quantities that are globally conserved, up to external sources.    In a typical theory, these will be charge, energy and momentum, and we will assume this to be the case for graphene as well.   
Hydrodynamics is a systematic way to truncate equations of motion for the local charge density $n(x)$,   energy density $\epsilon(x)$ and momentum density $\Pi_i(x)$, by treating the perturbative parameter as $l_{\mathrm{ee}} \partial_\mu$.    In fact, it is typical to instead study the dynamics of the thermodynamic conjugate variables: chemical potential $\mu(x)$, temperature $T(x)$ and relativistic velocity $u^\mu(x)$, respectively.  $u^\mu$ is subject to the usual constraint $u_\mu u^\mu = -v_{\mathrm{F}}^2$.

Note that throughout this paper, ``charge density" $n$ refers to the number density of electrons, minus the number density of holes:  $n=n_{\mathrm{elec}} - n_{\mathrm{hole}}$.   Thus, there are no factors of $e$ in the definition of $n$, or chemical potential $\mu$.\footnote{Therefore $[n] = [\text{length}]^{-d}$ and $[\mu]=[\text{energy}]$.}

The equations of motion of hydrodynamics are the local conservation laws, up to external sources.  We apply an external chemical potential $\mu_0$ via an external electromagnetic field $A_{\mathrm{ext}}^t = -\mu_0(\mathbf{x})/e$, $A^i_{\mathrm{ext}}=0$.   We employ relativistic notation with $v_{\mathrm{F}}=1$ temporarily.  The equations of hydrodynamics are\begin{subequations}\label{hydroeq}\begin{align}
\partial_\mu T^{\mu\nu} &= e F^{\mu\nu}_{\mathrm{ext}} J_\nu, \\
\partial_\mu J^\mu &= 0,
\end{align}\end{subequations}where $F^{ti}_{\mathrm{ext}} = -F^{it}_{\mathrm{ext}} = \partial_i \mu_0$ are the only non-vanishing components, $T^{\mu\nu}$ represents the expectation value of the local stress-energy density,  and $J^\mu$ the expectation value of the local charge density.     $T^{\mu\nu}$ and $J^\mu$ must be expressed in terms of $\mu$, $T$ and $u^\mu$ in order to obtain a closed set of equations.   One can show that there is a static solution to the hydrodynamic equations with $u^\mu = (1,0,0)$, $T=T_0=\text{constant}$, and $\mu(\mathbf{x})=\mu_0(\mathbf{x})$ \cite{lucas}.   Recall that $\mu_0(\mathbf{x})$ is slowly varying on the length scale $\xi$.  We will take this solution as the background state of our fluid.


Hydrodynamics is a perturbative expansion of (\ref{hydroeq}), where the perturbative expansion parameter is the number of derivatives of space and time.   At zeroth order, the equations of state are simply that $T^{\mu\nu}$ and $J^\mu$ are given by the  thermodynamic  relations we derived above:  \begin{subequations}\begin{align}
T^{\mu\nu} &=  (\epsilon + P)u^\mu u^\nu + P \eta^{\mu\nu}, \\
J^\mu &=  n u^\mu,
\end{align}\end{subequations} with $\epsilon$ the energy density and $P$ the pressure.   In the fluid's rest frame,  $T^{tt}=\epsilon$, $T^{ij} = P\mdelta^{ij}$, and $J^t=n$, with all other components vanishing.  At first order, \cite{hkms} showed that the most general first derivative corrections to $T^{\mu\nu}$ and $J^\mu$ consistent with symmetries and the local second law of thermodynamics are \begin{subequations}\begin{align}
T^{\mu\nu} &= (\epsilon +P )u^\mu u^\nu + P\eta^{\mu\nu}-2\mathcal{P}^{\mu\rho}\mathcal{P}^{\nu\sigma}\eta \partial_{(\rho} u_{\sigma)} - \mathcal{P}^{\mu\nu}\left(\zeta-\eta\right)\partial_\rho u^\rho   , \\
J^\mu &=  n u^\mu - \frac{\sigma_{\textsc{q}}}{e^2} \left(\eta^{\mu\nu}+u^\mu u^\nu\right)\left(\partial_\nu \mu - \frac{\mu}{T}\partial_\nu T - eF_{\nu\rho,\text{ext}}u^\rho\right),
\end{align}\end{subequations}
with $\eta,\zeta,\sigma_{\textsc{q}} >0$, and $\mathcal{P}^{\mu\nu} = \eta^{\mu\nu} + u^\mu u^\nu$.    Here $\eta$ and $\zeta$ are the shear and bulk viscosity respectively, and $\sigma_{\textsc{q}}$ is a ``quantum critical'' conductivity \cite{hkms}.
  Note that the external electromagnetic fields show up in the hydrodynamic gradient expansion in the charge current;  this happens because the charged fluid is sensitive only to the gradient in the total electrochemical potential \cite{pines}.  We allow for $P$, $n$, $\eta$, $\zeta$ and $\sigma_{\textsc{q}}$ to all be position-dependent, with their position dependence related to $\mu_{\mathrm{ext}}$, as we will describe shortly in more detail.   

It has long been appreciated \cite{hkms} that $\sigma_{\textsc{q}}$ plays a fundamental role in hydrodynamic transport near quantum critical points.   More recently \cite{dsz} argued that $\eta$ could play a role in transport.   We will carefully detail how $\eta$ affects transport in this paper, analytically and numerically.

In our extension of this theory to graphene, we will also allow for Coulomb interactions of the fluid to be substantial enough to enter the hydrodynamic equations.  However, this should only alter the equations of state, as well as add a further contribution to $F_{\mu\nu,\mathrm{ext}}$ \cite{muller1}, and we will detail this in the subsequent section.   The constraints imposed on hydrodynamics from local positivity of entropy production \cite{hkms} are unchanged in the presence of Coulomb interactions, which are entirely accounted for via a modified $F_{\mathrm{ext}}^{\mu\nu}$.

It is sufficient in our calculation of $\sigma$, $\alpha$ and $\kappa$ to simplify $T^{\mu\nu}$ and $J^\mu$ and retain only the terms linear in velocity.   One finds, in $d=2$: \begin{subequations}\label{smallvhyd}\begin{align}
T^{ti} &= (\epsilon+P) v^i, \\
T^{ij} &= P\mdelta^{ij} - \eta \left(\partial^i v^j + \partial^j v^i\right) - (\zeta-\eta)\mdelta^{ij} \partial_k v^k, \\
J^i &= nv^i - \frac{\sigma_{\textsc{q}}}{e^2} \left(\partial_i (\mu - \mu_0)  - \frac{\mu}{T}\partial_i T \right).
\end{align}
\end{subequations}
We stress the novel role of $\sigma_{\textsc{q}}$, a new dissipative transport coefficient in a relativistic fluid, without a direct analogue in the canonical non-relativistic fluid.   This term is related to the underlying thermally excited electron-hole plasma, and the fact that electrons and holes can move in opposite directions under an applied electric field, contributing a net electric current.\footnote{It is qualitatively similar to the bipolar diffusion effect \cite{goldsmid, yoshino} -- however, in hydrodynamics the quasiparticle lifetimes are limited by $\hbar/k_{\mathrm{B}}T$, whereas in the bipolar diffusion effect these lifetimes are parametrically long, as in a Fermi liquid.}  There is no microscopic thermal conductivity -- instead, all microscopic dissipation related to electric and thermal gradients is controlled by $\sigma_{\textsc{q}}$.  


\subsection{Hydrodynamic Theory of Transport}
We are now ready to detail our hydrodynamic calculation of the transport coefficients defined in Section \ref{sectransco}.   We place our fluid in a box of length $L$ in each direction, with periodic boundary conditions on the fluid in every direction.   We then apply a constant background  $E_i$ and $\zeta_i$.\footnote{The application of a constant $\zeta_j$ on a periodic space is the reason why we cannot talk about driving the system with a constant temperature gradient, since the temperature is a periodic function in space.  One can formally implement $\zeta_i$ through deformations of the spacetime metric and external gauge fields \cite{hartnollads}.}  The static solution above is no longer a solution to the hydrodynamic equations of motion, sourced by these gradients.   Now, we generically expect to excite both a spatial electric current $J^i$, and a spatial heat current \begin{equation}\label{heatdef}
Q^i \equiv T^{ti} - \mu J^i.
\end{equation}
We can expand out $J^i$ and $Q^i$ locally as a Taylor series in $E_i$ and $\zeta_i$.   The transport coefficients in Section \ref{sectransco} are defined by retaining only the linear terms in $E_i$ and $\zeta_i$, and spatially averaging over the local charge and heat currents.   It is  sufficient to perform a linear response calculation about our previously identified static solution:
\begin{subequations}\begin{align}
\mu &\approx \mu_0(\mathbf{x}) + \mdelta \mu(\mathbf{x}) \mathrm{e}^{-\mathrm{i}\omega t},  \\
T &\approx T_0 + \mdelta T(\mathbf{x}) \mathrm{e}^{-\mathrm{i}\omega t}, \\
u^t &\approx 1, \\
u^i &\approx \mdelta v^i (\mathbf{x}) \mathrm{e}^{-\mathrm{i}\omega t},
\end{align}\end{subequations}
and then solve the linearized hydrodynamic equations -- this is equivalent to only keeping terms linear in $E_i$ and $\zeta_i$ in the full solution.   For ease of notation, we drop the ``$\mdelta$" in  front of the linear response perturbations in the remainder of the paper, but one should keep in mind that $\mu(\mathbf{x})$, $T(\mathbf{x})$, and $v^i$ are henceforth perturbatively small quantities.

Following \cite{lucas}, the linearized hydrodynamic equations (\ref{hydroeq}) can be shown to take the following form:\footnote{In this equation, derivatives act on all fields to the right, so $\partial_x \eta \partial_x v^x$ should be read as $\partial_x (\eta \partial_x v^x)$.} \begin{align}
 &\left(\begin{array}{ccc} - e^{-2}\partial_i \sigma_{\textsc{q}} \partial_i  &\  e^{-2}T_0^{-1} \partial_i \mu_0\sigma_{\textsc{q}} \partial_i  &\ \partial_j n   \\ e^{-2} \partial_i \mu_0\sigma_{\textsc{q}} \partial_i  &\ -e^{-2} T^{-1}_0\partial_i \mu_0^2\sigma_{\textsc{q}} \partial_i  &\ T_0\partial_j s  \\ n\partial_i   &\ s\partial_i  &\  - \partial_i (\zeta-\eta)\partial_j  - \partial_i \eta \partial_j - \partial_j \eta \partial_i  \end{array}\right)  \left(\begin{array}{c} \mu \\ T \\ v_j\end{array}\right) \notag \\
&\;\;\;\;\;\;\;\;\;\;\;  = \left(\begin{array}{c} -e^{-1}\partial_i \sigma_{\textsc{q}} (E_i - \mu_0 \zeta_i/e) \\ e^{-1} \partial_i \sigma_{\textsc{q}}  \mu_0 (E_i - \mu_0 \zeta_i/e) \\ en E_i + T_0 s\zeta_i \end{array}\right)  \label{bigeq}
\end{align}
Here \begin{equation}
s = \frac{\epsilon+P - \mu_0 n}{T_0}
\end{equation}is the entropy density of the background fluid.   $s$ and $n$ are not independent, and are related by thermodynamic Maxwell relations:  see Appendix \ref{appthermo}.   We have also employed \begin{equation}
\partial_i P =  n \partial_i \mu + s\partial_i T. 
\end{equation}
In particular, $s$ and $n$ are position dependent functions whose position dependence is entirely determined by the local chemical potential:  $s(\mathbf{x}) = s(T_0,\mu_0(\mathbf{x}))$, and similarly for $n$, $\eta$, and all other coefficients in the hydrodynamic equations.   The proper boundary conditions to impose on $\mu$, $T$ and $v_j$ are periodicity.   This forms a well-posed elliptic partial differential equation and can be numerically solved: see Appendix \ref{appfin}.   Combining (\ref{smallvhyd}) and (\ref{heatdef}), along with $\mu$, $T$ and $v^i$ as found from solving the linear system (\ref{bigeq}), we obtain $J_i(E_j,\zeta_j)$ and $Q_i(E_j,\zeta_j)$.  Spatially averaging these quantities and employing (\ref{transeq}), we obtain $\sigma_{ij}$,$\alpha_{ij}$ and $\bar\kappa_{ij}$. 

We cannot exactly compute these transport coefficients in general.   However, one can prove \cite{lucas} that Onsager reciprocity holds.  This is a non-trivial consistency check on the validity of our approach.   Furthermore, there exist scaling symmetries combining re-scalings of $\mu$, $T$ and $v_i$, as well as the equations of state;  these are listed in Appendix \ref{apprescale}.   These are helpful when we fit this theory to the data of \cite{crossno}.   These scaling symmetries are also present in the theory of \cite{hkms}, with the exception of a further scaling symmetry which only affects the viscosity and the length scale of disorder in this theory.

In the limit where \begin{equation}
\mu_0 = \bar\mu_0 +  u \hat\mu(\mathbf{x}),  \label{pertlimit}
\end{equation} 
with $\hat\mu$ an O(1) function but $u \ll \bar\mu_0$, the transport coefficients may be perturbatively calculated analytically, and for $\mu \ll k_{\mathrm{B}}T$,  we find that \begin{subequations}\label{drudeeq}\begin{align}
\sigma &\approx \frac{e^2v_{\mathrm{F}}^2n^2\tau}{\epsilon+P}, \\
\alpha &\approx \frac{ev_{\mathrm{F}}^2ns\tau}{\epsilon+P}, \\
\bar\kappa &\approx \frac{v_{\mathrm{F}}^2Ts^2\tau}{\epsilon+P},
\end{align}\end{subequations} 
and we find an analytical expression for $\tau$ with the following approximate form near the Dirac point: \begin{equation}
\frac{1}{\tau} \approx \frac{v_{\mathrm{F}}^2 u^2}{2} \left(\frac{\partial n}{\partial \mu}\right)^2 \left[\frac{e^2 }{\sigma_{\textsc{q}}(\epsilon+P)} + \frac{\eta+\zeta}{\xi^2} \frac{4\mu^2}{(\epsilon+P)^3} \right].  \label{taumain}
\end{equation}  Details of this calculation and a more precise (and complicated) formula are given in Appendix \ref{appmom}.    The requirement that we are ``far" from the Dirac point is that $\sigma_{\textsc{q}} \ll e^2 v_{\mathrm{F}}^2n^2\tau/(\epsilon+P)$.   Everything in (\ref{taumain}) except for $u$ is evaluated in the clean fluid with $u\rightarrow 0$.    (\ref{taumain}) makes clear that if $\eta/\xi^2$ is large, the $n$ and $\mu$ dependence of $\tau$ is not negligible even when $\mu \ll  k_{\mathrm{B}}T$, and we will verify this in numerical simulations in Section \ref{secnum}.    The validity of (\ref{hkmseq}) for $\kappa$ is not guaranteed far from the Dirac point in this perturbative limit, but can often be quite good in practice, when the density dependence of all parameters is accounted for.     Combining (\ref{drudeeq}) and (\ref{taumain}), we obtain the relativistic analogue of the perturbative results of \cite{andreev}. 

Noting that $n\sim \mu$ as $\mu\rightarrow 0$,  careful study of (\ref{hkmseq}) shows that the Lorentzian form of $\kappa(n)$ is not altered by plugging in this hydrodynamic formula for $\tau$,   while the form of $\sigma(n)$ can be quite distinct,  with $\sigma(n)$ no longer growing quadratically at larger $n$.   This helps explain why in Figure \ref{mainfig}, (\ref{hkmseq}) gave a quantitatively good fit to $\kappa(n)$, but not to $\sigma(n)$.

 \section{The Dirac Fluid in Graphene}\label{secgraphene}
 The previous section developed a general theory for relativistic fluids.    It is often said that the Dirac fluid in graphene is a ``quantum critical" system in two spatial dimensions \cite{vafek, schmalian, muller3}, and exhibits behavior analogous to the quantum critical regime at finite temperatures above the superfluid-insulator transition, although technical differences arise.  Let us review elementary features of the quantum critical behavior of graphene, and argue that our formalism remains appropriate for transport computations.

Assuming that the electrons in graphene are non-interacting, standard band theory calculations on a honeycomb lattice in two spatial dimensions with nearest-neighbor hopping give two species of Dirac fermions with low-energy dispersion relation
\begin{equation}
\epsilon(\mathbf{q}) \approx \hbar v_{\mathrm{F}} |\mathbf{q}|,  \label{eq1}
\end{equation}
Convincing experimental evidence for these massless Dirac fermions was given in \cite{geim2005, kim2005}.  There is a quantum critical point between electron and hole Fermi liquids at zero temperature in graphene, as the chemical potential $\mu$ is tuned through the Dirac point,  $\mu=0$.   At (any experimentally accessible) finite temperature $T$, and at $\mu \ll T$, an effectively relativistic plasma of electrons and holes forms, interacting via a $1/r$ Coulomb potential.    The strength of these Coulomb interactions is captured by a dimensionless number $\alpha_0$ analogous to the fine structure constant: \begin{equation}
\alpha_0 = \frac{e^2}{4\mpi \epsilon_0 \epsilon_{\mathrm{r}} \hbar v_{\mathrm{F}}} \approx \frac{1}{137} \frac{c}{\epsilon_{\mathrm{r}} v_{\mathrm{F}}},
\end{equation}
where $\epsilon_{\mathrm{r}} \sim 4$ is a dielectric constant,  $c \approx 3\times 10^8$ m/s is the speed of light,  $v_{\mathrm{F}} \approx 1.1 \times 10^6$ m/s is the Fermi velocity in graphene and $e$ is the charge of the electron.    In experiments, $\alpha_0 \sim \mathrm{O}(1)$, and so unlike quantum electrodynamics ($\alpha_{\mathrm{QED}} \approx 1/137$),  interactions are \emph{strong}.  $v_{\mathrm{F}}$ plays the role of the speed of light in an effectively relativistic electron-hole plasma, and in an experimentally accessible regime which we describe below,  one can use relativistic hydrodynamics to model thermoelectric transport in graphene.   

The exception to the emergent Lorentz invariance is the photon-mediated Coulomb interactions, which are the standard $1/r$ interaction of three spatial dimensions.   Further, because $v_{\mathrm{F}} \sim c/300$,  the Coulomb interaction is essentially non-local and instantaneous in time.   Despite this,  graphene shares many features with a truly relativistic plasma with ``speed of light" $v_{\mathrm{F}}$, including a ``quantum critical" diffusive conductivity $\sigma_{\textsc{q}}$ \cite{muller2}.

Analogously to in quantum electrodynamics, $\alpha$ is a marginally irrelevant interaction, and so the effective coupling constant runs.   At temperature $T\rightarrow 0$, we should replace $\alpha_0$ with \cite{schmalian} \begin{equation}
\alpha_{\mathrm{eff}} = \alpha_0 \left(1+\frac{\alpha_0}{4}\log\frac{\Lambda}{T_0}\right)^{-1} \label{eq3}
\end{equation}
where $\Lambda \sim 8.4\times 10^4$ K is a cutoff related to the graphene band structure (the energy scale at which the dispersion is no longer linear).  Note that although the running of $\alpha_{\mathrm{eff}}$ causes a logarithmically increasing velocity $v_{\mathrm{F}}$ in (\ref{eq1}), when we write $v_{\mathrm{F}}$ in this paper, we are always referring to the bare velocity, $1.1\times 10^6$ m/s.    

At the experimentally accessible temperatures ($T_0\sim 70$ K) where the plasma described above is most likely not suppressed by local disorder in $\mu$ \cite{crossno},  (\ref{eq3}) gives $\alpha_{\mathrm{eff}} \sim 0.25$.    And so the experiments likely probe the dynamics of a strongly interacting quasi-relativistic plasma.   It is such a regime where hydrodynamics is a good approximation.   More carefully, the electron-electron scattering length has quantum critical scaling \cite{muller2} \begin{equation}
l_{\mathrm{ee}} \sim \frac{\hbar v_{\mathrm{F}}}{\alpha_{\mathrm{eff}}^2 k_{\mathrm{B}}T_0} \sim 100 \; \mathrm{nm},  \label{leeeq}
\end{equation}
where we have plugged in for experimentally reasonable values of the parameters.   Indeed, pump-probe experiments provide evidence that the electron-electron interaction time,  $l_{\mathrm{ee}}/v_{\mathrm{F}} \sim 10^{-13} \; \mathrm{s}$, is consistent with (\ref{leeeq}) \cite{breusing, johannsen}.   Furthermore, it is believed that the dominant source of disorder in graphene are charge puddles, which are fluctuations in the local charge density.   It is now possible to find samples of graphene where these fluctuations are correlated on the length scale (\ref{leeeq}) \cite{xue}.   In these cleanest samples, the experimental evidence thus points to the validity of a hydrodynamic description, such as the one we advocate in this paper.

Most computations of the thermodynamic and hydrodynamic coefficients in graphene are based on kinetic theory, which requires a quasiparticle description to be sensible, and so are valid as $\alpha_{\mathrm{eff}} \rightarrow 0$ ($T_0\rightarrow 0$), when the plasma becomes weakly interacting.   However, the experiments are likely not in this weakly interacting regime, and $\log \alpha_{\mathrm{eff}}$ corrections to these properties are not negligible.   As such, we will allow \emph{all} coefficients in the equations of motion to be fit parameters.    We will also neglect the fact that the running of $\alpha_{\mathrm{eff}}(T_0)$ allows for certain thermodynamic relations for a strictly scale invariant, relativistic fluid to be violated.   This assumption is justified in Appendix \ref{apprun}.


We must also take into account the long range Coulomb interactions in our hydrodynamic description.   This can be done following \cite{muller1}.   The Coulomb potential introduces a local electric field and must be included in $F_{\mathrm{ext}}^{\mu\nu}$: \begin{equation}
A^t_{\mathrm{ext}} = \mu_{\mathrm{ext}}-\varphi = \mu_{\mathrm{ext}} - \varphi_{\mathrm{ext}} - \mdelta \varphi   \label{eq21}
\end{equation}
where \begin{equation}
\varphi(\mathbf{x}) = \int \mathrm{d}^2\mathbf{y}\; K(\mathbf{x}-\mathbf{y}) \; n(\mathbf{y}),
\end{equation}
with $K$ a Coulomb kernel whose specific form \cite{sarma2009} is not necessary for our purposes, and the $n$ the charge density.    At $T_0=0$,   $K(r) = \alpha_{\mathrm{eff}}/r$;  at finite $T_0$, this is cut-off at long wavelengths due to thermal screening \cite{sarma2009}.    In (\ref{eq21}) we have separated the effects of Coulomb screening into two contributions:  $\varphi_{\mathrm{ext}}$, which alters the background disorder profile, so that $\mu_0 \ne \mu_{\mathrm{ext}}$, and $\mdelta\varphi$, which is the infinitesimal Coulomb potential created by the change in charge density $\mdelta n$, proportional to $E_i$ and $\zeta_i$. 

The time-independent equations of motion depend only on $ T$, $ v_i$,  the sources $ E_i$ and $ \zeta_i$, and the electrochemical potential\begin{equation}
\mdelta \Phi \equiv \mdelta \mu +  \mdelta \varphi .
\end{equation}
This is a direct consequence of the tightly constrained way that $F_{\mathrm{ext}}$ and $\mu$ enter the hydrodynamic gradient expansion.    If we solve for $\mdelta\Phi$ instead of $\mdelta\mu$,  we find that Coulomb screening does not affect dc transport at all:  more precisely, the equations of motion are identical to those in Section \ref{sechydro}, but with $\mdelta \Phi$ replacing $\mdelta \mu$.   That dc transport is insensitive to Coulomb screening of the hydrodynamic degrees of freedom was also noted in \cite{muller1} in a homogeneous fluid by appealing to the random phase approximation.\footnote{See also the discussion in \cite{pines, lucasMM}.}    It is therefore appropriate to directly apply the formalism of Section \ref{sechydro} to study dc thermal and electric transport in the Dirac fluid in graphene.   To maintain notation with Section \ref{sechydro},  we will continue to  refer to $\Phi$ as $\mu$ in our linear response theory, with the understanding that this includes corrections due to Coulomb screening.

\section{Numerical Results}\label{secnum}
Having argued that the theory of Section \ref{sechydro} is an acceptable approximation for dc transport in the Dirac fluid in graphene, we now present the results of numerical simulations of (\ref{bigeq}).   In our numerics, we assume that the equations of state of the graphene fluid are as follows: \begin{subequations}\label{numericmain}\begin{align}
n(\mu_0) &= \left(\frac{k_{\mathrm{B}}T_0}{\hbar v_{\mathrm{F}}}\right)^2\left[C_2 \frac{\mu_0}{k_{\mathrm{B}}T_0} + C_4 \left(\frac{\mu_0}{k_{\mathrm{B}}T_0}\right)^3\right], \\
s(\mu_0) &= \frac{k_{\mathrm{B}}^3T_0^2}{(\hbar v_{\mathrm{F}})^2}\left[C_0 + \frac{C_2}{2} \left(\frac{\mu_0}{k_{\mathrm{B}}T_0}\right)^2 -\frac{C_4}{4} \left(\frac{\mu_0}{k_{\mathrm{B}}T_0}\right)^4\right], \\
\eta(\mu_0) &= \frac{(k_{\mathrm{B}}T_0)^2}{\hbar v_{\mathrm{F}}^2} \eta_0, \\
\zeta(\mu_0) &= 0, \\
\sigma(\mu_0) &= \frac{e^2}{\hbar}\sigma_0,
\end{align}\end{subequations}
with $C_{0,2,4}$, $\sigma_0$ and $\eta_0$ dimensionless constants.   The form of $n$ and $s$ are consistent with thermodynamic Maxwell relations -- see Appendix \ref{appthermo}.   We take the disorder profile to be random sums of sine waves, and normalize the disorder distribution so that \begin{equation}
\left\langle (\mu_0 - \bar\mu_0)^2\right\rangle = u_0^2 (k_{\mathrm{B}}T_0)^2.  \label{numericmain2}
\end{equation}
The shortest wavelength sine wave in the problem is taken to have wavelength $\xi=l_{\mathrm{ee}}$ in all of our numerics.   There is an exact symmetry of the problem under which $\xi$ can be made arbitrary, so long as we rescale $\eta$ and $\zeta$ by a factor of $(\xi/l_{\mathrm{ee}})^2$ -- see Appendix \ref{apprescale}.   We have chosen this value of $\xi$ as it is roughly consistent with previous experimental observations \cite{xue}, and also the smallest value for which a hydrodynamic description is sensible.    More details on numerical methods are in Appendix \ref{appfin}.

\begin{figure}[t]
\centering
\includegraphics[width=7in]{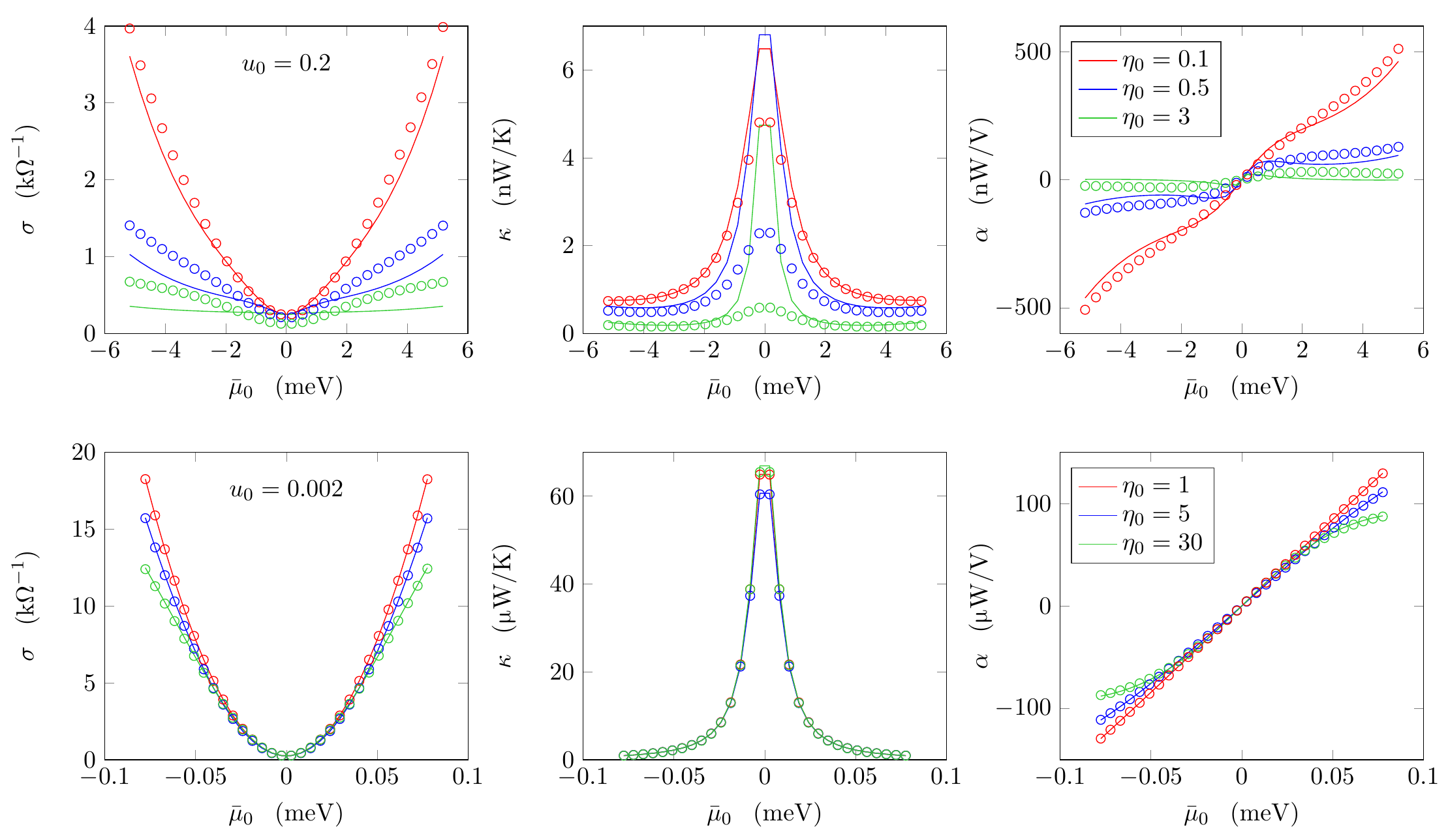}
\caption{Numerical computations of transport coefficients with $C_1=C_2=\sigma_0=1$ and $C_4=0$.   The top row has $u_0=0.2$, and the bottom row has $u_0=0.002$.   Solid lines are our theoretical results (using the particular disorder realizations studied) and the circular markers are numerical results.   Averages are taken over 20 disorder realizations.  $T_0=75$ K and we employ the value of $v_{\mathrm{F}}$ in graphene.  }
\label{viscfig}
\end{figure}

An example of our numerical results is shown in Figure \ref{viscfig}, where the results of varying the dimensionless viscosity $\eta_0$ are shown.    When the charge puddle sizes are $\sim 20$ K, as in experiment \cite{xue}, the  value of $\eta_0$ dramatically alters the transport coefficients as a function of density.   In particular, the $\sigma(n)$ and $\alpha(n)$ curves are substantially flattened, an effect which is predicted using (\ref{taumain}).   Further, the peak in $\kappa(n)$ is substantially smaller than predicted perturbatively, and $\kappa(n)$ does not shrink to 0 as $n\rightarrow\infty$, as predicted in \cite{hkms}.   In contrast, in a limit of extremely weak disorder (temperature at which the Dirac fluid  emerges $\sim$ 0.2 K),  the transport coefficients are relatively insensitive to the viscosity (assuming that $\eta_0/C_0 \sim 1$, as expected for a strongly interacting quantum fluid).  

\begin{figure}[t]
\centering
\includegraphics[width=7in]{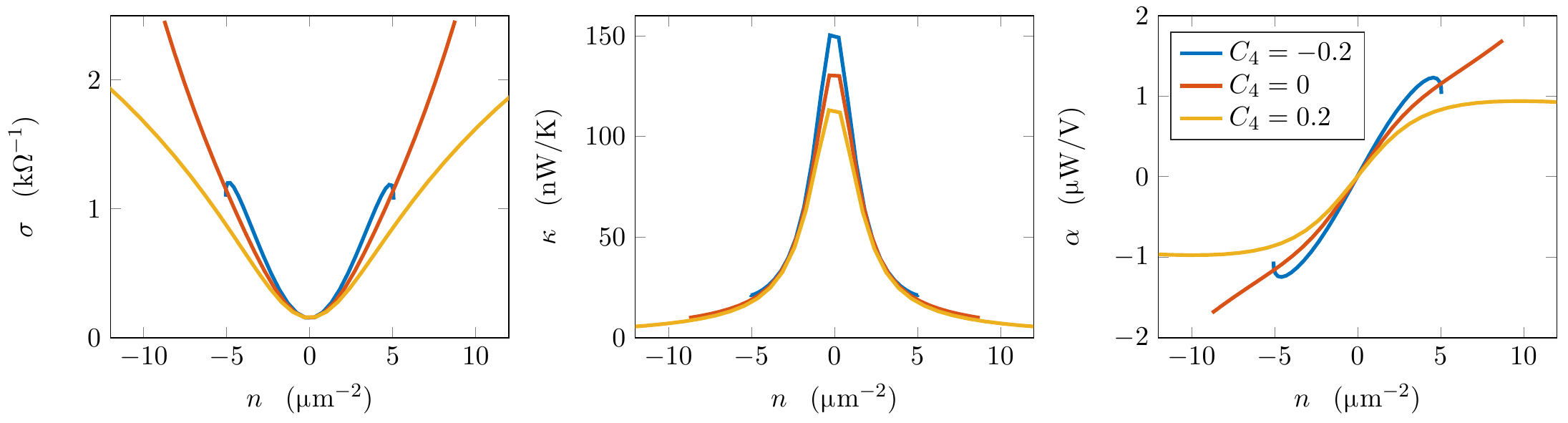}
\caption{Numerical computations of transport coefficients with varying $C_4$, $C_1=\sigma_0=1$, $\eta_0=3$, $C_2=0.2$ and $u_0=0.2$.    The sharp change in behavior when $C_4<0$ is a consequence of $n(\mu)$ not being monotonically increasing at large $\mu$.    Averages are taken over 20 disorder realizations.  $T_0=75$ K and we employ the value of $v_{\mathrm{F}}$ in graphene.  }
\label{c4fig}
\end{figure}

We also show the consequences of a non-zero $C_4$ in Figure \ref{c4fig}.   The most important effect of $C_4$ is that $n$ and $\bar\mu_0$ are no longer proportional -- in particular, when $C_4>0$ we see that at larger $n$ both $\sigma$ and $\alpha$ decrease much more slowly with $n$.   Whenever $C_4\ne 0$, the equations of state become badly behaved at large $\mu$,  because $s(\mu)$ or $n(\mu)$ becomes a non-monotonically increasing function.  At lower temperatures ($T\lesssim 50$ K) in Figure \ref{mainfigT}, this begins to be an issue in the codes for the equations of state  we use to compare to experiment.  This implies that higher order terms in the equations of state (associated with more fit parameters) are necessary.

\subsection{Comparison to Experiment}

We now describe in more details the lessons to be drawn from our fit to experimental data, shown in Figures \ref{mainfig} and \ref{mainfigT}.   Due to a total of 6 fit parameters (3 which determine the overall scales in the plots, and 3 which alter the shapes of curves),  we did not perform an exhaustive analysis and find a statistically optimal fit.   We found that most choices of parameters do not agree well with data, and the fit we have presented serves as a proof of principle that hydrodynamics can explain many important features of the experiment \cite{crossno}, as we now discuss.


 To obtain data at lower temperatures, we have taken disorder realizations from $T_0=75$ K, using our standard assumption $\xi=l_{\mathrm{ee}}(T_0=75\; \mathrm{K})$, and simply lowered the temperature.  We also keep $u_0T_0$ constant as a function of $T_0$.  Formally, this implies that at lower temperatures $\xi<l_{\mathrm{ee}}$, as $l_{\mathrm{ee}}\sim T^{-1}$;  this may be problematic for the validity of hydrodynamics.    A conservative solution, employing the rescaling symmetries of our theory, is to simply double $\xi$,  and quadruple $\eta_0$:  all data is exactly identical, except that for all data points taken in Figure \ref{mainfigT},  $\xi>l_{\mathrm{ee}}$ and $\eta_0$ increases.

\begin{figure}[t]
\centering
\includegraphics[width=3.5in]{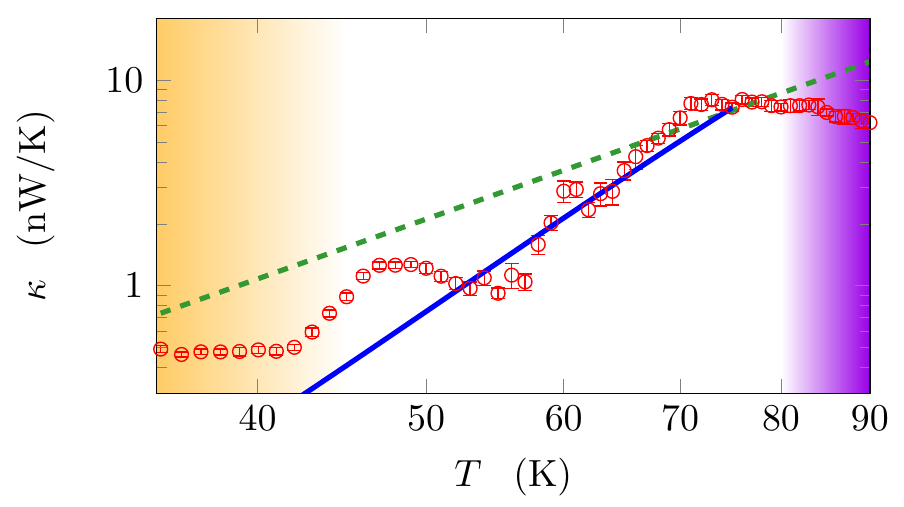}
\caption{A comparison of our numerical computation of $\kappa(T)$ with experimental results of \cite{crossno}  at the charge neutrality point ($n=0$).   The red data points are experimental data from \cite{crossno}, the blue curve is our disorder-averaged simulation (using identical parameters to Figure \ref{mainfigT}), and the green dashed curve is the perturbative prediction $\kappa \sim T^3$ for comparison.  Data is shown on a log-log scale.  The yellow shaded region denotes where Fermi liquid behavior is observed; the purple shaded region denotes the likely onset of electron-phonon coupling.   }
\label{Tscalefig}
\end{figure}

\begin{figure}[t]
\centering
\includegraphics[width=3.5in]{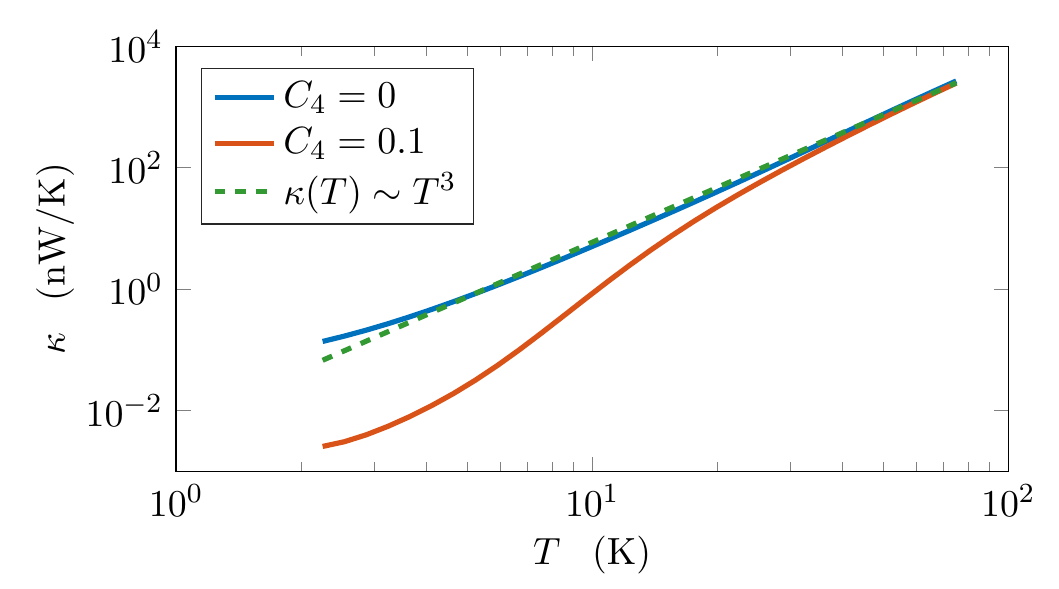}
\caption{A comparison of $\kappa(T)$ in simulations with varying $C_4$.  We take $C_0=1$, $C_2=\eta_0=0.1$, $\sigma_0=1$, $u_0=0.1$ (at $T_0=75$ K).   At large $T$ both scenarios have $\kappa \sim T^3$;  at lower $T$ the fluid with $C_4>0$ undergoes a dramatic drop in $\kappa(T)$, similar to that observed in experiment.}
\label{Tscaletheoryfig}
\end{figure}

Figure \ref{Tscalefig} revisits the $T$-dependence in $\kappa$.  Assuming that disorder is weak, we employ (\ref{drudeeq}) and (\ref{taumain}) to determine the scaling of $\kappa$:  since  $s\sim T^2$, $\epsilon+P\sim T^3$, $\partial n/\partial \mu \sim T$, and the viscosity dependence in $\tau$ is negligible, we obtain $\tau \sim T$ and $\kappa \sim T^3$.   That numerics and experiment are not consistent with this power law is a sign of the strong non-perturbative effects, and suggests that observing  power law signatures of hydrodynamics may only be possible in the cleanest samples:
see Figure \ref{Tscalefig}.    Figure \ref{Tscaletheoryfig} suggests that the sharp dependence in $T$ observed in experiment is a consequence of $C_4>0$ and is not a robust scaling regime.\footnote{For this particular simulation, the disorder becomes large enough at $T\lesssim 7.5$ K that disorder realizations with $C_4=0.1$ sometimes have unphysical thermodynamic behavior.}  As noted in \cite{crossno}, this dramatic $T$-dependence of $\kappa$, in contrast with the very weak $T$-dependence of $\sigma$, at the Dirac point, is a tell-tale sign of hydrodynamics that is not captured by competing theories, such as the bipolar diffusion effect.

The fits to $\sigma(n)$ and $\sigma(T)$ are not as good as the fits to $\kappa$.  Nonetheless, our theory does help to explain the slow growth in $\sigma$ away from the Dirac point, as a consequence of a fluid with both non-negligible viscosity and large disorder, as in Figure \ref{viscfig}.   Our simulations also correctly predict that the conductivity is an increasing function of $T$,  an entirely non-perturbative effect, in Figure \ref{mainfigT}.   This is at odds with predictions from kinetic theory in the Dirac fluid, which predict that $\sigma(n=0,T) \sim \alpha^{-2}_{\mathrm{eff}}$ should be decreasing with $T$ due to the $T$-dependence in $\alpha_{\mathrm{eff}}$ \cite{muller2}.   Any residual contact resistance \cite{wang13} will also increase the growth rate of $\sigma(n)$ away from the Dirac point, and as such will be closer fit by our numerical results in Figure \ref{mainfig}.

The most surprising thing about the fit is the  large values of all coefficients, compared to previous theories.   For example, it is predicted \cite{vafek, schmalian} that $C_0 \lesssim 3.4$, and we find $C_0 \sim 10$.   This is a direct consequence of the large values of the density $n$ over which the Dirac fluid is present (as measured by where strong deviations from the Wiedemann-Franz law occur).  The naive theoretical estimate is that the Dirac fluid should not extend past about $n\sim 40 \; \mmu \mathrm{m}^{-2}$,\footnote{We have used theoretically predicted values of $C_0$ and $C_2$ \cite{schmalian}, and assumed that the Dirac fluid ends when the $\mu$-dependent contribution to $s$ is comparable to the $T$-dependent contribution.} yet we see the Dirac fluid all the way to about $n\sim 200 \; \mmu\mathrm{m}^{-2}$; we will comment more on this issue shortly.    As in non-relativistic fluid dynamics, our hydrodynamic theory has a large number of rescaling symmetries (Appendix \ref{apprescale}), and these rescaling symmetries turn out to lead to very large values for all hydrodynamic coefficients as a consequence of the large scale on the density axis in Figure \ref{mainfig}.

Another consequence of this rescaling is a dramatically large shear viscosity:  $\eta_0\sim 100$.   It is now canonical to normalize this by the entropy density, and so the ``proper units" to measure $\eta$ in are $\eta_0/C_0\approx 10$.   This scaling is a consequence of a proposition \cite{kss} that strongly coupled theories would have $\eta/s \approx \hbar/4\mpi k_{\mathrm{B}}$, or $\eta_0/C_0 \approx 1/4\mpi$.  The viscosity is a helpful measure of the interaction strength in a theory;  if the interactions are perturbatively controlled by a small parameter $g$, then we expect $\eta \sim g^{-2}$;  only when the interaction strength is large can $\eta \sim s$, up to a prefactor of order unity.  Hence, coming close to saturating the bound of \cite{kss} is a signature that the fluid is strongly interacting. Our estimate for $\eta_0/C_0$ is about 100 times larger than the bound of \cite{kss}.   Smaller values of $\eta/s \sim 0.5 \hbar /k_{\mathrm{B}}$ have been reported in other experiments in cold atomic gases \cite{cao} and quark-gluon plasma \cite{luzum}.   The possibility of adding the Dirac fluid to a list of strongly interacting quantum fluids is tantalizing, and a more direct measure of $\eta$ in the Dirac fluid is of interest. 

One possibility is that our bare coefficients $C_0$, $\eta_0$ etc. are anomalously large because \cite{crossno} has measured the average charge density in the entire sample.   However, some regions of the sample (notably close to the contacts on the edges \cite{huard}, or regions very close to impurities) may have such large local values of $\mu_0$ that they are always in a Fermi liquid regime.  So long as these Fermi liquid regimes do not percolate across the entire sample, our hydrodynamic description of transport may be quite reasonable in the bulk.  However, these regions have a much smaller compressibility, and so can absorb a lot of charge relative to a clean Dirac fluid.    It may be that the total averaged charge density is then not equal to the averaged hydrodynamic charge density,  leading us to overestimate $n$.   Rescaling $n \rightarrow \lambda n$ would rescale $C_0 \rightarrow \lambda C_0$ and $\eta_0 \rightarrow \lambda^2 \eta_0$.   Choosing $\lambda = 0.2$, in accordance with our previous estimates on the regime of the Dirac fluid at $T_0\sim 75$ K, we obtain $C_0\sim 2$ and $\eta_0/C_0 \sim 2$, which are both reasonable for a strongly interacting fluid.   

As noted previously, we expect that future measurements in cleaner samples may give a wider separation between $l_{\mathrm{ee}}$ and $\xi$.   Together with a better understanding of edge effects and the charge puddle profile, we expect this approach to lead to cleaner estimates of $C_{0,2,4}$, $\eta_0$ and $\sigma_0$. 

\section{Phonons in Graphene}\label{secphonon}
Throughout this work we have neglected the effects of electron-phonon coupling in graphene \cite{fong, fong2}.   In this section, we provide some brief qualitative comments on the role of electron-phonon coupling in the experiment \cite{crossno}, and discuss signatures for future experiments.

Generically, phonons extract both energy and momentum from the electronic fluid, and in doing so hamper a hydrodynamic description.\footnote{The hydrodynamic description of transport reduces to a diffusion equation for the conserved electrical current.   Historically, this was modeled via resistor networks \cite{kirkpatrick, derrida}.}  In graphene, the acoustic branch(es) of phonons have dispersion relation \cite{hwang07} \begin{equation}
\omega_{\mathrm{ac}}(\mathbf{q}) \approx \hbar v_{\mathrm{a}} |\mathbf{q}|
\end{equation} with  $v_{\mathrm{a}}\approx 2\times 10^4 $ m/s and so $v_{\mathrm{a}} \ll v_{\mathrm{F}}$.   By considering conservation of energy and momentum in electron-phonon scattering events, one finds that the phonon energies are negligible, and thus the scattering event can be treated as elastic from the point of view of the electrons.

If only acoustic phonons couple to the electronic fluid, we may approximate that the momentum conservation equation is modified, following the phenomenology of \cite{hkms}: \begin{equation}
\partial_\mu T^{\mu i} \approx F^{\nu i}_{\mathrm{ext}} J_\nu - \frac{T^{ti}}{\tau_{\mathrm{a}}} .
\end{equation}
The latter term implies that the momentum of the electronic fluid degrades at a constant rate $\tau^{-1}_{\mathrm{a}}$, which we take to be \begin{equation}
\frac{1}{\tau_{\mathrm{a}}} = \mathcal{B} T^a,
\end{equation}where $a>0$ and $\mathcal{B}>0$ are constants that are phenomenological.  \cite{hwang07} computed their values using kinetic theory and found $a=4$ far from the Dirac point.   This effect has been observed experimentally \cite{efetov},  but $a$ is expected to change near the Dirac point.   Following arguments similar to \cite{ashcroft, hwang07}, we can estimate $a$ by assuming a quasiparticle description of transport, and that the dominant events are absorption or emission of a single phonon.   Since acoustic phonons cannot effectively carry away energy, a Dirac quasiparticle of energy $\epsilon$ can scatter into $\sim \epsilon$ states.   All phonons with relevant momenta are thermally populated, and we estimate the scattering rate to be proportional to the momentum of the phonon.   Thus we estimate, using that the typical quasiparticle energy is $\epsilon \sim T$, $a=1+1=2$.\footnote{As there is no large Fermi surface with $\mu \gg k_{\mathrm{B}}T$, no further corrections to $a$ are necessary, as in usual metals.}

\begin{figure}[t]
\centering
\includegraphics[width=3.5in]{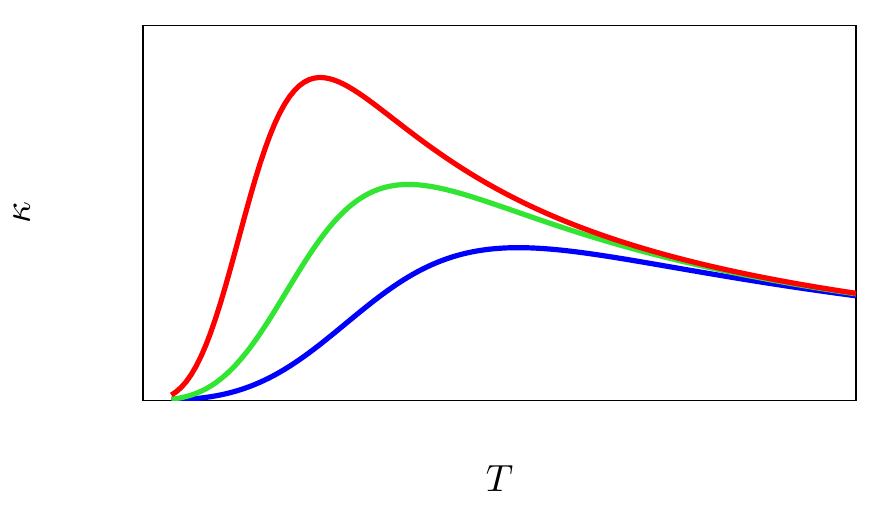}
\caption{A sketch of $\kappa(T)$, accounting for coupling to acoustic phonons, for samples of graphene with three different amounts of disorder (measured by $u$).  We take $a=3$, $b=1$ in this plot.}
\label{ephfig}
\end{figure}

Assuming that the charge puddles are small and can be accounted for perturbatively,  $\kappa$ is approximately given by (\ref{hkmseq}) at the Dirac point, with \begin{equation}
\frac{1}{\tau} \approx  \frac{u^2}{2\sigma_{\textsc{q}} (\epsilon+P)}\left(\frac{\partial n}{\partial \mu}\right)^2 +  \mathcal{B}T^a = \frac{\mathcal{A} u^2}{T^b} + \mathcal{B}T^a.
\end{equation}
Our analytic theory predicts $b=1$.  The contribution $\kappa$ from electron-phonon coupling is negligible so long as \begin{equation}
T \ll  T^* \equiv  \left(\frac{\mathcal{A}}{\mathcal{B}}u^2\right)^{\frac{1}{b+a}}.
\end{equation}
Note that $T^*(u)$ is an increasing function -- the weaker the disorder, the lower the temperature at which electron-phonon coupling cannot be neglected in the Dirac fluid.   The thermal conductivity scales as \begin{equation}
\kappa \sim \left\lbrace\begin{array}{ll}  T^{2+b} &\  T\ll T^* \\ T^{2-a} &\ T\gg T^*\end{array}\right.
\end{equation}
If $a>2$, we find phenomenology quite similar to that observed in \cite{crossno}, with $\kappa(T)$ growing non-monotonically.   We also find that \begin{equation}
\kappa(T^*) = \mathcal{C} (T^*)^{2-a},
\end{equation}
a result which can be tested in experiment by measuring $T^*$ via the peak in $\kappa$ for different samples.  The prefactor of this proportionality $\mathcal{C}$ may not be very sensitive to the particular sample, since it is independent of $u$.   Figure \ref{ephfig} shows a sketch of $\kappa(T)$, accounting for acoustic phonons, in three samples with different disorder strengths $u$.   This mechanism is also consistent with the fact that the cleaner samples in \cite{crossno} had peaks in $\kappa(T)$ at lower temperatures, which suggests our proposed mechanism for the non-monotonicity in $\kappa(T)$ is sensible.   Although our perturbative quasiparticle-based argument found $a=2$ above, the presence of local charge puddles may increase the effective value of $a$ to somewhere between 2 and 4, and lead to $\kappa(T^*)$ be a decreasing function.   A careful analysis of electron-phonon coupling in disordered Dirac fluids is worth more study.   

At higher temperatures, we expect optical phonons to couple non-negligibly to the electron fluid.   These phonons can exchange both energy and momentum effectively,  and at this point we expect the measured thermal conductivity to increase due to electron-phonon coupling.   In \cite{crossno},  there is a sharp upturn in $\kappa(T)$ at all densities at temperatures of 100 K, which is likely due to activation of optical phonons in the boron nitride subtrate \cite{crossno2}.

\section{Conclusions}

We have developed a theory of transport in realistic hydrodynamic electron fluids near a quantum critical point.    This theory provides a substantially improved quantitative fit to $\kappa(n)$ and $\sigma(n)$ \emph{simultaneously}.    We have further found reasonable quantitative fits to $\sigma(T)$ and $\kappa(T)$ at the Dirac point, giving us valuable information about the mechanism of momentum relaxation beyond the theory of \cite{hkms}.

  Although we have managed to find fluids where the growth in $\sigma(n)$ is quite slow, there are still differences between the shape of $\sigma(n)$ found numerically and in experiment.   There are numerous possibilities for residual discrepancies.   One of the most important possibilities is that the disorder is so strong that the full thermodynamic equation of state is necessary -- in this paper, we have only kept the three leading order terms.   Alternatively, we may simply not have found the correct equation of state of graphene.  A disorder profile more subtle than superimposed sines and cosines may also be responsible for deviations with our theory, although our investigation into this possibility suggests that other disorder profiles cause $\sigma(n)$ to have more substantial density dependence.    We have assumed that the disorder profile is unaltered both by changes in $T$ and in $\bar\mu_0$.   This is a very strong assumption and need not be true.   Finally, there may be other sources of momentum relaxation, such as out-of-plane distortions in the graphene lattice, or interactions with phonons.   An understanding of the aforementioned issues is an important future task, though may be quite challenging given the possibility that strong interactions in the Dirac fluid at $T\sim 70$ K may lead to the failure of standard perturbative techniques.   The most fruitful direction for resolving at least some of these questions may be directly in experiments: for example, techniques to directly resolve the local charge density on length scales $\lesssim$ 10 nm are well known \cite{xue}, and can shed light into the evolution of $\mu_0$ as a function of $T$, as well as the spatial correlations in $\mu_0$.
  
Experimentally, it may be possible to generate samples of graphene with much weaker charge puddles using suspended devices \cite{bolotin, mayorov}.  Thermodynamic measurements can also be used to determine the coefficients $C_{0,2,4}$, though these measurements are complicated by the presence of disorder, as we discuss in Appendix \ref{appthermo}.   Nonetheless, measurements of the specific heat and compressibility in the Dirac fluid will serve as valuable guideposts for future hydrodynamic models.   Such measurements have been made in the Fermi liquid \cite{yacoby2007}, and their extension to the Dirac fluid form the basis for worthwhile experiments.

Previous experiments which measured the ac conductivity \cite{horng} were not in the hydrodynamic limit.   Comparing the momentum relaxation time $\tau$ between measurements of $\kappa$, and a putative Drude peak in ac transport, may provide a quantitative test of our theory.   Studying magnetotransport \cite{hkms} may also be a fruitful direction in experiments.  A theoretical discussion of transport at finite frequency and magnetic field beyond the weak disorder limit will appear elsewhere \cite{lucas2}.   The thermopower  of graphene has recently been measured at $T\sim 200$ K \cite{ghahari},  and it would be interesting to measure $\sigma$, $\alpha$ and $\kappa$ in the same sample in the Dirac fluid and compare with our hydrodynamic formalism.

\addcontentsline{toc}{section}{Acknowledgements}
\section*{Acknowledgements}
We would like to thank Richard Davison and Koenraad Schalm for helpful discussions.

A.L. and S.S. are supported by the NSF under Grant DMR-1360789 and MURI grant W911NF-14-1-0003 from ARO.
Research at Perimeter Institute is supported by the Government of Canada through Industry Canada 
and by the Province of Ontario through the Ministry of Research and Innovation.  J.C. thanks the support of the FAME Center, sponsored by SRC MARCO and DARPA.   K.C.F. acknowledges Raytheon BBN Technologies' support for this work.  P.K. acknowledges partial support from the Gordon and Betty Moore Foundation's EPiQS Initiative through Grant GBMF4543 and Nano Material Technology Development Program through the National Research Foundation of Korea (2012M3A7B4049966).  

\begin{appendix}

\section{Thermodynamics}\label{appthermo}
In this appendix and in every subsequent appendix, we will work in units where $\hbar=k_{\mathrm{B}}=v_{\mathrm{F}}=e=1$.     It is straightforward using dimensional analysis to restore these units.  

We consider the equations of state of the relativistic plasma in a relativistic strongly interacting fluid in $d=2$.    Without specific microscopic details, these are very general facts about relativistic plasmas without an intrinsic mass scale (or gap).  The discussion generalizes straightforwardly to other $d$.    The only relevant energy scales are the temperature $T$ and the chemical potential $\mu$.     We have the general Gibbs-Duhem relation:\begin{equation} \label{duhem}
\epsilon + P = \mu n + Ts,
\end{equation}where $\epsilon$ is the energy density, $P$ is the pressure, $s$ is the entropy density and $n$ is the charge density ($n=0$ at the particle-hole symmetric Dirac point).   In a relativistic fluid, \begin{equation}
P  = T^3 \mathcal{F}\left(\frac{\mu}{T}\right)
\end{equation}for some dimensionless function $\mathcal{F}$.  Thermodynamic identities imply that \begin{subequations} \label{qmu}\begin{align}
n &= \frac{\partial P}{\partial \mu} = T^2 \mathcal{F}^\prime \left(\frac{\mu}{T}\right) \\
s &= \frac{\partial P}{\partial T} = 3T^2 \mathcal{F} -  \mu T\mathcal{F}^\prime \left(\frac{\mu}{T}\right) = \frac{3P - \mu n }{T}.
\end{align}\end{subequations}Combining (\ref{duhem}) and (\ref{qmu}) we obtain \begin{equation}
\epsilon = 2P.
\end{equation}

The hydrodynamic description is only sensible for $\mu \ll T$ -- for $\mu \gg T$ the standard Fermi liquid description applies.   Furthermore, the equations of state of the Dirac fluid are charge conjugation symmetric, implying that $\mathcal{F}$ is an even function of $\mu$.   So we Taylor expand: \begin{equation}
\mathcal{F}\left(\frac{\mu}{T}\right) \approx \frac{C_0}{3} + \frac{C_2}{2} \left(\frac{\mu}{T}\right)^2 +  \frac{C_4}{4} \left(\frac{\mu}{T}\right)^4.  \label{feq}
\end{equation} Using (\ref{qmu}): \begin{subequations}\begin{align}
n &= C_2 \mu T + C_4 \frac{\mu^3}{T}, \\
s &= C_0T^2 + \frac{C_2 \mu^2}{2} - \frac{C_4 \mu^4}{4T^2} .
\end{align}\end{subequations}
We also require that $C_0,C_2\ge 0$, so that $s\ge 0$ and that $n/\mu$ is positive as $\mu\rightarrow 0$, as it should be.


\subsection{Thermodynamics of Disordered Fluids}
Already at this point we can make interesting predictions about the thermodynamics of the strongly interacting hydrodynamic regime in graphene.   For concreteness, let us suppose that the background chemical potential is \begin{equation}
\mu_0(\mathbf{x}) = \bar\mu_0 + \hat\mu(\mathbf{x}),
\end{equation}
with $\bar\mu_0$ a constant and $\hat\mu$ a zero-mean random function;  for simplicity, suppose that $\hat\mu$ is evenly distributed about zero, and has a disorder correlation length of $\xi\gtrsim l_{\mathrm{ee}}$, so that the hydrodynamic description applies.   In this case, spatially averaging over $\mu_0$, we find \begin{equation}
\langle \epsilon\rangle = \frac{2C_0}{3}T^3  + C_2T\left(\bar\mu_0^2+\left\langle \hat\mu^2\right\rangle \right) + \frac{C_4}{2}\left(\bar\mu_0^4 + 6\bar\mu_0^2\left\langle \hat\mu^2\right\rangle + \left\langle \hat\mu^4\right\rangle\right) + \cdots.
\end{equation}
The $\cdots$ denotes higher order terms in the equation of state that we have neglected.  A similar expression can be found for the charge density: \begin{equation}
\langle n\rangle = C_2 T \bar\mu_0 + \frac{3C_4}{T}\bar\mu_0 \left\langle \hat\mu^2\right\rangle + \cdots.
\end{equation}

Let us focus on a clean limit where $\hat\mu$ is very small relative to $T$.    Let us also assume that we are close to the Dirac point, so that only $C_0$ and $C_2$ terms need to be kept.   Thermodynamics then gives tight constraints on the behavior of measurable quantities: specific heat and compressibility,  in an experimentally testable regime, due to the ability to easily tune both $T$ and $\bar\mu_0$ (the average charge density) experimentally.  In the limit above, the (spatially averaged) compressibility $\mathcal{K}$ is \begin{equation}
\frac{1}{\mathcal{K}} = \frac{\partial \langle n\rangle}{\partial \mu}  = C_2T.
\end{equation}
where as before, we use $\langle\cdots\rangle$ to denote a uniform spatial average.    Note that the independence of $\mathcal{K}$ to $\bar\mu_0$ and $\hat\mu$ is simply a consequence of the fact that we did not expand (\ref{feq}) to quartic order.  The spatially averaged specific heat \begin{equation}
 c = \frac{\partial \langle \epsilon\rangle}{\partial T} = 2C_0T^2 + C_2\left(\bar\mu_0^2+\left\langle \hat\mu^2\right\rangle\right).
\end{equation}

The experimental consequence of this result is as follows.   Very close to the Dirac point, we expect that $\mathcal{K}$ is approximately constant.   Restoring all dimensional prefactors, we can therefore set \begin{equation}
C_2 \approx \frac{(\hbar v_{\mathrm{F}})^2}{\mathcal{K}k_{\mathrm{B}}T}
\end{equation}
and re-write \begin{equation}
c \approx 2C_0 \frac{k_{\mathrm{B}}^3T^2}{(\hbar v_{\mathrm{F}})^2} + 2\frac{\bar\mu_0^2+\left\langle \hat\mu^2\right\rangle}{\mathcal{K}T} \approx 2C_0 \frac{k_{\mathrm{B}}^3T^2}{(\hbar v_{\mathrm{F}})^2} + 2\frac{\left\langle \hat\mu^2\right\rangle}{\mathcal{K}T} + \frac{2\mathcal{K}  n^2}{T} .
\end{equation}
We thus see that the quadratic dependence in $c(n)$ gives us an independent measurement of $\mathcal{K}$ through a measurement of the heat capacity.   In principle, a quadratic polynomial fit to $c(n)$ thus determines both $\mathcal{K}$ and $C_0$,  up to the residual effects of disorder, which will lead to an overestimate of $C_0$.   Repeating measurements of $\mathcal{K}$ directly, as well as $c(n)$ at different $T$,  provide non-trivial checks on the above theory.   It is important to note that this argument does \emph{not} rely on the validity of hydrodynamics, only that graphene is gapless, $\bar\mu_0 \ll T$, and that $\hat\mu$ is very small.   Of these three requirements, the last poses the biggest experimental hurdle.

In the above argument, there is no reason a priori why to truncate the Taylor expansion to neglect $C_4$ and higher order corrections, beyond appealing to the weak disorder limit.   In particular, inclusion of $C_4$ complicates our ability to obtain an accurate measure of $\bar\mu_0$ from $n$.    The argument above is simply meant to give a flavor for the constraints on measurable quantities imposed by scale invariant thermodynamics.   A more systematic treatment is likely necessary to make quantitative contact with future experiments.

\section{Rescaling Symmetries of dc Transport}\label{apprescale}
Solutions to (\ref{bigeq}) are invariant, up to global rescalings, under certain rescalings of the linearized hydrodynamic equations of motion.   These symmetries are, assuming $\lambda>0$ is a constant scaling parameter:  
\begin{subequations}\begin{align}
\eta &\rightarrow \lambda^2 \eta, \;\; x \rightarrow \lambda x; \\
\eta &\rightarrow \lambda \eta, \;\; \sigma \rightarrow \lambda \sigma, \;\; \alpha \rightarrow \lambda\alpha, \;\; \kappa \rightarrow \lambda\kappa, \;\; P \rightarrow \lambda P; \\
\alpha &\rightarrow \lambda \alpha, \;\; \kappa \rightarrow \lambda^{2}\kappa, \;\; \eta \rightarrow \lambda^{-2} \eta, \;\;\; \mu_0 \rightarrow \lambda \mu_0, \;\; C_2 \rightarrow \lambda^{-2}C_2,\;\; C_4 \rightarrow \lambda^{-4}C_4,\;\; \mathrm{etc.}; \\
\alpha &\rightarrow \lambda \alpha, \;\; \kappa \rightarrow \lambda\kappa, \;\; \sigma \rightarrow \lambda\sigma, \;\; \eta \rightarrow \lambda^{-1} \eta.
\end{align}\end{subequations}
Everything not listed is invariant.  $\zeta$ and $\eta$ have the same scalings, as do $\sigma$ and $\sigma_{\textsc{q}}$, and so we have only listed some of these parameters.     

These rescalings are useful to help us compare simulations to experimental data from graphene.  The latter three rescalings can be used to help fix the overall magnitude of $\kappa$ and $\sigma$, as well as the values of $n$, as measured in experiment.    These are exactly analogous to the symmetries of the Navier-Stokes equation, which allow us to reduce all such hydrodynamic problems to a universal equation, up to a single dimensionless parameter.   \cite{crossno} neither measured the viscosity directly nor the charge puddle size, and the first scaling above implies that we cannot determine viscosity alone.  So, as mentioned in the main text, we assume that $\xi = l_{\mathrm{ee}}$,  the shortest possibile value of $\xi$ for which hydrodynamics seems sensible.   

\section{Weak Disorder}\label{appmom}
Many analytic results can be obtained in the limit where the disorder strength is ``small".   We provide detailed derivations of all such results in this appendix.    We introduce disorder as in (\ref{pertlimit}).    Below we denote $n_0 = n(\bar\mu_0)$, etc.

The perturbative solution is found exactly as was done in \cite{lucas}:   we split the velocity field into a constant piece $\bar v_i \sim u^{-2}$, and a fluctuating zero-mean piece $\hat v_i \sim u^{-1}$;  similarly, $ \mu \sim  T \sim u^{-1}$.   It proves convenient to work in Fourier space.   At $\mathrm{O}(u^{-1})$, the momentum conservation equation becomes \begin{equation}
-\mathrm{i}k_i \left(n_0  \mu(\mathbf{k})+s_0 T(\mathbf{k})\right) = \eta_0 k^2  \hat v_i(\mathbf{k}) + \zeta_0 k_ik_j  \hat v_j(\mathbf{k}),
\end{equation} 
and the conservation laws become (at the same order) \begin{subequations}\begin{align}
0 &= \mathrm{i}k_i \left(\hat n(\mathbf{k})  \bar v_i + n_0  \hat v_i(\mathbf{k})\right)+ \sigma_{\textsc{q}0} k^2 \left( \mu(\mathbf{k}) - \frac{\mu_0}{T_0} T(\mathbf{k})\right), \\
0 &= \mathrm{i}k_i T_0 \left(\hat s(\mathbf{k})  \bar v_i + s_0  \hat v_i(\mathbf{k})\right) -\mu_0\sigma_{\textsc{q}0} k^2 \left( \mu(\mathbf{k}) - \frac{\mu_0}{T_0} T(\mathbf{k})\right).
\end{align}\end{subequations}
Combining these equations we obtain expressions for $ T$, $ \mu$ and $k_i  \hat v_i$: \begin{subequations}\begin{align}
k_i  \hat v_i(\mathbf{k}) &= -\frac{\mu_0 \hat n(\mathbf{k}) + T_0\hat s(\mathbf{k})}{\mu_0 n_0 + T_0s_0} k_i  \bar v_i, \\
 T(\mathbf{k}) &= -\frac{\mathrm{i}k_i  \bar v_i}{\sigma_{\textsc{q}0}k^2(\epsilon_0+P_0)^2}\left(\sigma_{\textsc{q}0}k^2 (\eta_0+\zeta_0)(\mu_0 \hat n + T_0\hat s)T_0 - T_0n_0(T_0s_0\hat n-T_0n_0\hat s)\right), \\
 \mu(\mathbf{k}) &= -\frac{\mathrm{i}k_i  \bar v_i}{\sigma_{\textsc{q}0}k^2(\epsilon_0+P_0)^2}\left(\sigma_{\textsc{q}0}k^2 (\eta_0+\zeta_0)(\mu_0 \hat n + T_0\hat s)\mu_0 + T_0s_0(T_0s_0\hat n-T_0n_0\hat s)\right).
\end{align}\end{subequations}
Spatially averaging over the momentum conservation equation at $\mathrm{O}(u^0)$, and defining: 
\begin{equation}
(\epsilon+P)\tau^{-1}_{ij}  \bar v_j \equiv  \sum_{\mathbf{k}} \mathrm{i}k_i \left[\hat n(-\mathbf{k})  \mu(\mathbf{k}) + \hat s(-\mathbf{k})  T(\mathbf{k})\right],
\end{equation}
we find that \begin{equation}
\tau^{-1}_{ij} = \sum_{\mathbf{k}} \frac{k_ik_j}{k^2} \frac{\left| T_0s_0 \hat n(\mathbf{k}) - T_0n_0 \hat s(\mathbf{k})\right|^2 + \sigma_{\textsc{q}0}k^2 (\eta_0+\zeta_0) \left|\mu_0 \hat n(\mathbf{k}) + T_0\hat s(\mathbf{k})\right|^2}{\sigma_{\textsc{q}0}(\epsilon_0+P_0)^3}   \label{taueq1}
\end{equation}and that the spatially averaged momentum equation reduces to \begin{equation}
0  = n_0  E_i + T_0s_0  \zeta_i -(\epsilon_0+P_0)\tau^{-1}_{ij} \bar v_j .  \label{hydrocpeq}
\end{equation}
In this equation, we have used the fact that $ J_i \approx n \bar v_i$ at leading order in perturbation theory.   The resulting transport coefficients are analogous to (\ref{drudeeq}): \begin{subequations}\begin{align}
\sigma_{ij} &= \frac{n_0^2}{\epsilon_0+P_0}  \tau_{ij}, \\
\bar\alpha_{ij} = \alpha_{ij} &=  \frac{n_0s_0}{\epsilon_0+P_0}\tau_{ij}, \\
\bar\kappa_{ij} &= \frac{Ts_0^2}{\epsilon_0+P_0}\tau_{ij}.
\end{align}\end{subequations}In the expression for $\sigma$, we have not included a $\sigma_{\textsc{q}0}$ contribution, as was done in \cite{hkms}, as this is a subleading order in perturbation theory.

Using our Taylor expanded equations of state for the fluid and assuming $C_4=0$, since \begin{equation}
\hat s(\mathbf{k}) \approx C_2 \mu_0 \hat\mu(\mathbf{k}) = \frac{\mu_0}{T_0}\hat n(\mathbf{k}),
\end{equation}we can simplify (\ref{taueq1}) to \begin{equation}
\tau^{-1}_{ij} = \sum_{\mathbf{k}} \frac{k_ik_j}{k^2} \frac{\left(T_0s_0-\mu_0n_0\right)^2 +4 \sigma_{\textsc{q}0}k^2 (\eta_0+\zeta_0) \mu_0^2 }{\sigma_{\textsc{q}0}(\epsilon_0+P_0)^3}\left|\hat n(\mathbf{k})\right|^2
\end{equation}
Similar results were presented (in a different format) in \cite{dsz}, though the practical consequences of this formula, as discussed in the main text, have not previously been understood.

We cannot take the naive limit where $\sigma_{\textsc{q}0}\sim u^{-2}$ in order to recover (\ref{hkmseq}) in full generality.     The simplest way to see that something goes wrong is to study $\bar\kappa$ near $\bar\mu_0=0$ (more precisely, $\bar\mu_0\sim u$):  if $\sigma_{\textsc{q}} \sim u^{-2}$  we find that $\tau \sim u^{-4}$, and this implies that the heat current (and thus $\bar\kappa$) would be parametrically larger than anticipated.   Thus our perturbative scaling breaks down.   The breakdown of the perturbative theory for $u\sim \bar\mu_0$ was also advocated in \cite{lucas}.  
 
 Although we have argued there are problems in principle with (\ref{hkmseq}) when $\bar\mu_0 \sim u$, even when $u$ is perturbatively small, in practice  the mean-field model of \cite{hkms} can be quite good in practice near $\bar\mu_0\sim u$, as shown in Figure \ref{viscfig}.   Note that it is also important that $C_0 T^2 \gg C_2 u^2$ -- when this limit is violated, we see substantial deviations from (\ref{hkmseq}) at all $\bar\mu_0$, as shown in Figure \ref{viscfig}.  This may be a consequence of our assumption that $\sigma_{\textsc{q}}$ is independent of $\mu_0$.

\section{Equations of State of the Dirac Fluid}\label{apprun}
The thermodynamics of graphene is similar to that presented in Appendix \ref{appthermo}, with some minor differences.    Perturbative computations and renormalization group arguments, valid as $T\rightarrow 0$, give \cite{vafek, schmalian} \begin{subequations}\label{cb}\begin{align}
C_0 &= \frac{9\mzeta(3)}{\mpi} \left(\frac{\alpha_{\mathrm{eff}}}{\alpha_0}\right)^2 \approx 3.44 \left(\frac{\alpha}{\alpha_0}\right)^2, \\
C_2 &= \frac{4\log 2}{\mpi}  \left(\frac{\alpha_{\mathrm{eff}}}{\alpha_0}\right)^2 \approx 0.88 \left(\frac{\alpha}{\alpha_0}\right)^2.
\end{align}\end{subequations}
(\ref{cb}) can be derived by computing the thermodynamics of 2 species of non-interacting Dirac fermions, with Coulomb interactions leading to a logarithmically increasing Fermi velocity \cite{vafek, schmalian}.    As $\alpha_{\mathrm{eff}}(T)$ is not a constant, this implies that the entropy has an additional contribution related to the logarithmic $T$ dependence of $C_{0,2}(\alpha_{\mathrm{eff}})$.   Assuming $C_0$ and $C_2$ above, and assuming $C_4=0$ for simplicity as its value for free fermions is quite small \cite{muller1}, we find: \begin{equation}
   s = C_0 T^2 \left[1 + \frac{\alpha_{\mathrm{eff}}}{6}\right]+ \frac{C_2}{2}\mu^2\left[1 + \frac{\alpha_{\mathrm{eff}}}{2}\right]  \label{saeff}
   \end{equation}
This equation directly implies $\epsilon>2P$.   Using the estimate $\alpha_{\mathrm{eff}}\sim 0.25$ from above, we see that the corrections to $s$ (and $\epsilon$) are rather minor ($<10$\%);  $n$ is unchanged.   In fact,  (\ref{saeff}) is not quite right: the computation of (\ref{cb}) is only a leading order perturbative computation:  there are corrections to (\ref{cb}) at higher orders in $\log \alpha_{\mathrm{eff}}$.    Nonetheless, our general conclusion that $\epsilon>2P$ is possible in graphene continues to hold, given any logarithmic running of the Fermi velocity.

As noted above, these theoretical computations of the thermodynamic coefficients in graphene are all perturbative computations in $\alpha_{\mathrm{eff}}$,  yet we only expect $\alpha_{\mathrm{eff}}/\alpha_0\sim 0.5$:  there is no reason to expect that higher order corrections, which can be as large as $\sim \log \alpha_{\mathrm{eff}}$, are negligible.  More sensitive experiments may find discrepancies with Lorentz invariant hydrodynamics, associated with these peculiar properties of the Dirac fluid.   Similar logarithms can appear in $\sigma_{\textsc{q}}$ \cite{muller2} and $\eta$ \cite{muller3}, and in both cases, for the reasons above, we neglect these logarithms and use the theory of Section \ref{sechydro}.

\section{Numerical Methods}\label{appfin}
We solved the hydrodynamic equations (\ref{bigeq}) on a periodic domain of size $L\times L$, employing pseudospectral methods \cite{trefethen} using a basis of $N$ Fourier modes in each direction, with $25 \lesssim N \lesssim 43$.   For simplicity, we set $T_0=1$, as this can be restored straightforwardly by dimensional analysis.   Our numerical methods involves approximating continuous partial differential equations in the form \begin{equation}
\mathsf{L} \mathbf{u} = \mathbf{s}.
\end{equation}
 $\mathbf{u}$ contains the linear response fields $\mu$, $T$, $v_x$ and $v_y$, evaluated on a uniformly distributed discrete grid, and $\mathbf{s}$ contains the source terms, linear in $E_i$ and $\zeta_i$, evaluated at the same points.   $\mathsf{L}$ is a matrix with two zero eigenvectors, which correspond to constant shifts in $\mu$ and $T$ respectively.  We thus remove two rows of $\mathsf{L}$ and replace them with constraints that $\mu(\mathbf{0}) = T(\mathbf{0}) = 0$.  A simple matrix inversion thus gives $\mathbf{u} = \mathsf{L}^{-1}\mathbf{s}$.   Inverting this $(4N^2-2)\times (4N^2-2)$ matrix four times (once for sources $E_{x,y}$ and $\zeta_{x,y}$) limits the size of the domain we can analyze.   More complicated algorithms exist \cite{domaindec} to solve such problems but we did not find finite size effects to qualitatively alter our comparison to experimental data, as we discuss below.
 
As mentioned in the main text, our disorder realizations consisted of random sums of sine waves.  More precisely, \begin{equation}
\mu_0(\mathbf{x}) = \bar\mu_0 + \sum_{|n_x|,|n_y| \le k} \hat\mu_0(n_x,n_y) \sin\left(\phi_x + \frac{2n_x\mpi x}{k\xi}\right)\sin\left(\phi_y + \frac{2n_y\mpi y}{k\xi}\right)
\end{equation} 
with $\bar\mu_0$ a constant, $\hat\mu_0(n_x,n_y)$ uniformly distributed on $[-c,c]$ where $c=\sqrt{(2-\mdelta_{n_x,0}-\mdelta_{n_y,0})/2}$, and $\phi_{x,y}$ uniformly distributed on $[0,2\mpi)$.   The lack of heavy tails in $\hat\mu_0(n_x,n_y)$, perhaps associated with point-like impurities, is consistent with experiment \cite{yacoby2007}.  The form of $c$ is chosen so that we do not add random charge density bias to our disorder (as the zero mode has no amplitude), and so that all Fourier modes included at finite wave number have the same average amplitude.

\subsection{Finite Size Effects}
\begin{figure}[t]
\centering
\includegraphics[width=7in]{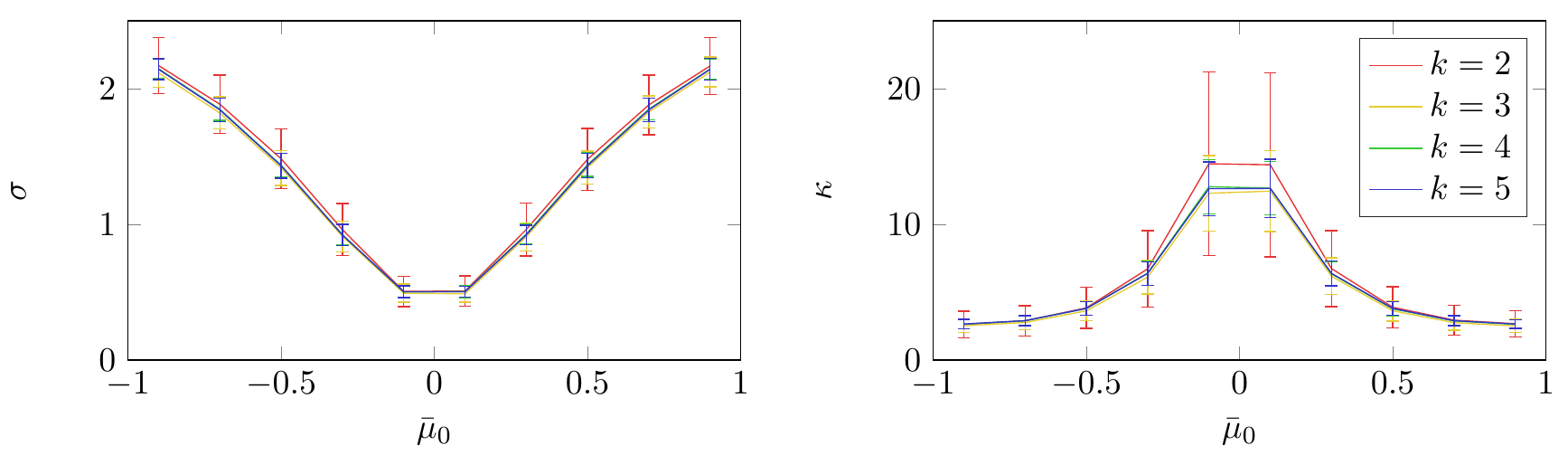}
\caption{Finite size effects with $u_0=0.3$, $C_0=3$, $C_2=1$, $\sigma_0=1$, $\eta_0=20$.  Numerical averages are performed over 100 disorder realizations.}
\label{finitesize2plot}
\end{figure}

\begin{figure}[t]
\centering
\includegraphics[width=7in]{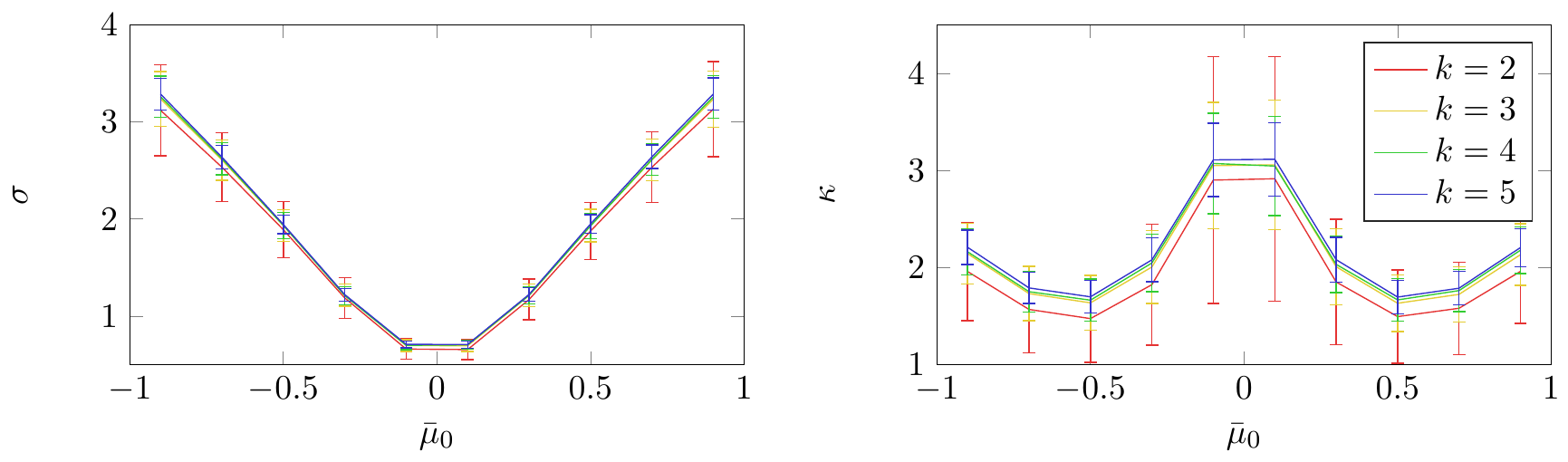}
\caption{Finite size effects with $u_0=0.3$, $C_0=1$, $C_2=1$, $\sigma_0=1$, $\eta_0=1$.  Numerical averages are performed over 100 disorder realizations.}
\label{finitesizeplot}
\end{figure}

The first source of finite size effects is simply related to the fact that we only have a finite number of disorder modes.   Averaging over a large number of ensemble samples allows us to approximately, but not exactly, undo this effect:  see Figures \ref{finitesize2plot} and  \ref{finitesizeplot}.   In both cases,  we used $8k+3$ grid points in each direction for various $k$.   To the best of our knowledge, in all numerical simulations we have studied, it appears as though the result converges to a finite fixed answer as $k\rightarrow \infty$.  However, residual error from finite size effects may lead to some error in our estimation of the thermodynamic and hydrodynamic coefficients of the Dirac fluid in graphene.

\begin{figure}[t]
\centering
\includegraphics[width=3.3in]{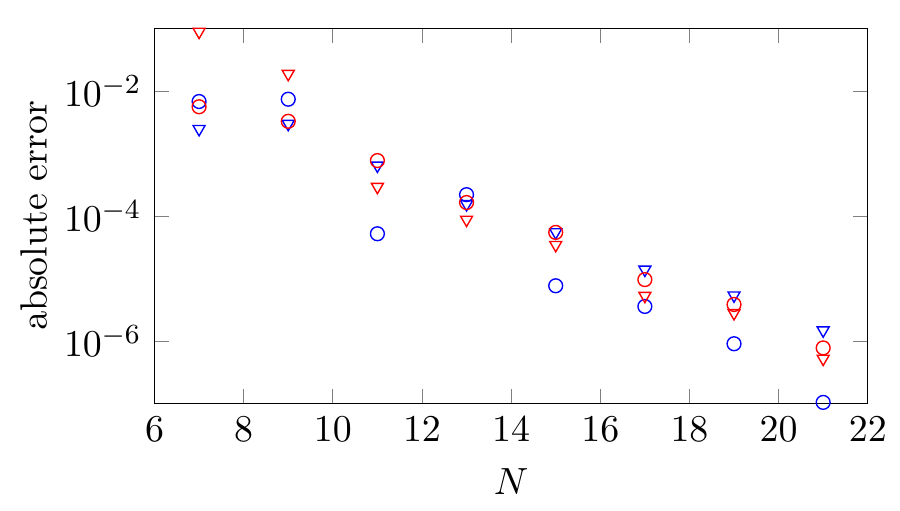}
\caption{Exponential convergence of our pseudospectral code with an increasing number of grid points.  We computed $\kappa$ and $\sigma$ using our  code, employing the ``experimental" equations of state given in Figure \ref{mainfig},  and the disorder profile $\mu_0(x) = \bar\mu_0 + 2u\cos(2\mpi x/L)\cos(2\mpi y/L)$.   Red data points denote the error in $\kappa$, and blue points denote the error in $\sigma$.   Circles denote data at $\bar\mu_0 = 2u$, and triangles at $\bar\mu_0 = 0.4u$.  Absolute error is determined by (e.g.) $|\sigma(N)-\sigma(29)|/\sigma(29)$,  where we use the data points at $N=29$ as a reference point.}
\label{spectralfig}
\end{figure}

The other source of finite size effects is related to the finite number of grid points in our pseudospectral methods.   However, we expect standard exponential accuracy \cite{trefethen} in the number of grid points per $\xi$, which we have taken to be at least 10 in all figures in the main text.   Numerical evidence suggests that our spectral methods have converged to within about 0.1--1\% of the correct answer by this relatively small number of grid points per $\xi$, depending on the precise equations of state used.  In the case of the experimentally relevant parameters used in Figure \ref{mainfig}, we see exponential convergence of our spectral methods with increasing grid points, with numerical error of only 0.1\% by the time the number of grid points per $\xi$ is 11, as shown in Figure \ref{spectralfig}.     This spectral convergence is dramatically faster in the weak disorder limit.

Methods are known to improve our simple algorithms, which can reduce both types of finite size effects discussed above.   Given the preliminary nature of the experiments to which we compare our simulations, we have found the numerical errors described above tolerable.

\subsection{Dimensional Analysis}
We have performed numerical computations in dimensionless units,   since we can trivially restore the units to our numerical results via dimensional analysis.   Setting $\hbar=k_{\mathrm{B}}=e=v_{\mathrm{F}}=T_0=1$ completely non-dimensionalizes the problem, while setting no dimensionless parameters to unity.    We can now trivially restore the units as follows:  \begin{subequations}\begin{align}
L &= \frac{\hbar v_{\mathrm{F}}}{k_{\mathrm{B}}T_0} \times L_{\mathrm{numerics}} \sim (100 \; \mathrm{nm}) \times L_{\mathrm{numerics}} \\
\mu &= k_{\mathrm{B}}T_0  \times \mu_{\mathrm{numerics}} \sim (5\; \mathrm{meV}) \times \mu_{\mathrm{numerics}}, \\
\sigma &= \frac{e^2}{\hbar}  \times \sigma_{\mathrm{numerics}} \sim (0.25\; \mathrm{k\Omega}^{-1}) \times \sigma_{\mathrm{numerics}}, \\
\alpha &= \frac{k_{\mathrm{B}}e}{\hbar}  \times \alpha_{\mathrm{numerics}} \sim \left(20\; \frac{\mathrm{nW}}{\mathrm{V}}\right) \times \alpha_{\mathrm{numerics}}, \\
\kappa &= \frac{k_{\mathrm{B}}^2T_0}{\hbar}  \times \kappa_{\mathrm{numerics}} \sim \left(0.1\; \frac{\mathrm{nW}}{\mathrm{K}}\right)  \times \kappa_{\mathrm{numerics}}.
\end{align}\end{subequations}
We have also noted the approximate scale of each important physical quantity in the problem for convenience.


  \end{appendix}

\addcontentsline{toc}{section}{References}
\bibliographystyle{unsrt}
\bibliography{theorypaperbib}

\begin{thebibliography}{10}

\bibitem{pines}
D.~Pines and P.~Nozi\`eres.
\newblock \emph{The Theory of Quantum Liquids, Volume I}
  \href{http://www.amazon.com/Theory-Of-Quantum-Liquids-Advanced/dp/0201407744}{(W.
  A. Benjamin, 1966)}.

\bibitem{ashcroft}
N.~W. Ashcroft and N.~D. Mermin.
\newblock \emph{Solid-State Physics}
  \href{http://www.amazon.com/Solid-State-Physics-Neil-Ashcroft/dp/0030839939/ref=sr_1_1?ie=UTF8&qid=1439625478&sr=8-1&keywords=ashcroft+and+mermin}{(Brooks
  Cole, 1976)}.

\bibitem{kadanoff}
L.~P. Kadanoff and P.~C. Martin.
\newblock ``Hydrodynamic equations and correlation functions",
  \href{http://www.sciencedirect.com/science/article/pii/0003491663900782}{\textsl{Annals
  of Physics} \textbf{24} 419 (1963)}.

\bibitem{molenkamp}
M.~J.~M. de~Jong and L.~W. Molenkamp.
\newblock ``Hydrodynamic electron flow in high-mobility wires",
  \href{http://journals.aps.org/prb/abstract/10.1103/PhysRevB.51.13389}{\textsl{Physical
  Review} \textbf{B51} 11389 (1995)},
  \href{http://arxiv.org/abs/cond-mat/9411067}{\texttt{arXiv:cond-mat/9411067}}.

\bibitem{weber}
C.~P. Weber, N.~Gedik, J.~E. Moore, J.~Orenstein, J.~Stephens, and D.~D.
  Awschalom.
\newblock ``Observation of spin Coulomb drag in a two-dimensional electron
  gas",
  \href{http://www.nature.com/nature/journal/v437/n7063/full/nature04206.html}{\textsl{Nature}
  \textbf{437} 1330 (2005)}.

\bibitem{lilly}
L.~Yang, J.~D. Koralek, J.~Orenstein, D.~R. Tibbetts, J.~L. Reno, and M.~P.
  Lilly.
\newblock ``Doppler velocimetry of spin propagation in a two-dimensional
  electron gas",
  \href{http://www.nature.com/nphys/journal/v8/n2/abs/nphys2157.html}{\textsl{Nature
  Physics} \textbf{8} 153 (2012)}.

\bibitem{andreev}
A.~V. Andreev, S.~A. Kivelson, and B.~Spivak.
\newblock ``Hydrodynamic description of transport in strongly correlated
  electron systems",
  \href{http://journals.aps.org/prl/abstract/10.1103/PhysRevLett.106.256804}{\textsl{Physical
  Review Letters} \textbf{106} \texttt{256804} (2011)},
  \href{http://arxiv.org/abs/1011.3068}{\texttt{arXiv:1011.3068}}.

\bibitem{succiturb}
M.~Mendoza, H.~J. Herrmann, and S.~Succi.
\newblock ``Preturbulent regimes in graphene flow",
  \href{http://journals.aps.org/prl/abstract/10.1103/PhysRevLett.106.156601}{\textsl{Physical
  Review Letters} \textbf{106} \texttt{156601} (2011)}.

\bibitem{tomadin}
A.~Tomadin, G.~Vignale, and M.~Polini.
\newblock ``A Corbino disk viscometer for 2d quantum electron liquids",
  \href{http://journals.aps.org/prl/abstract/10.1103/PhysRevLett.113.235901}{\textsl{Physical
  Review Letters} \textbf{113} \texttt{235901} (2014)},
  \href{http://arxiv.org/abs/1401.0938}{\texttt{arXiv:1401.0938}}.

\bibitem{vignale}
A.~Principi and G.~Vignale.
\newblock ``Violation of the Wiedemann-Franz law in hydrodynamic electron
  liquids",
  \href{http://journals.aps.org/prl/abstract/10.1103/PhysRevLett.115.056603}{\textsl{Physical
  Review Letters} \textbf{115} \texttt{056603} (2015)}.

\bibitem{polini}
I.~Torre, A.~Tomadin, A.~K. Geim, and M.~Polini.
\newblock ``Non-local transport and the hydrodynamic shear viscosity in
  graphene", \href{http://arxiv.org/abs/1508.00363}{\texttt{arXiv:1508.00363}}.

\bibitem{levitovhydro}
L.~Levitov and G.~Falkovich.
\newblock ``Electron viscosity, current vortices and negative nonlocal
  resistance in graphene",
  \href{http://arxiv.org/abs/1508.00836}{\texttt{arXiv:1508.00836}}.

\bibitem{bandurin}
D.~A.~Bandurin \emph{et al.}
\newblock ``Negative local resistance due to viscous electron backflow in
  graphene", \href{http://arxiv.org/abs/1509.04165}{\texttt{arXiv:1509.04165}}.

\bibitem{mackenzie}
P.~J.~W. Moll, P.~Kushwaha, N.~Nandi, B.~Schmidt, and A.~P. Mackenzie.
\newblock ``Evidence for hydrodynamic electron flow in $\mathrm{PdCoO}_2$",
  \href{http://arxiv.org/abs/1509.05691}{\texttt{arXiv:1509.05691}}.

\bibitem{vandermarel}
D.~van~der Marel, H.~J.~A. Molegraaf, J.~Zaanen, Z.~Nussinov, F.~Carbone,
  A.~Damascelli, H.~Eisaki, M.~Greven, P.~H. Kes, and M.~Li.
\newblock ``Quantum critical behaviour in a high-$T_{\mathrm{c}}$
  superconductor",
  \href{http://www.nature.com/nature/journal/v425/n6955/full/nature01978.html}{\textsl{Nature}
  \textbf{425} 271 (2003)}.

\bibitem{hussey}
R.~A.~Cooper \emph{et al.}
\newblock ``Anomalous criticality in the electrical resistivity of
  $\mathrm{La}_{2-x}\mathrm{Sr}_x\mathrm{CuO}_4$",
  \href{http://www.nature.com/nature/journal/v425/n6955/full/nature01978.html}{\textsl{Science}
  \textbf{323} 603 (2009)}.

\bibitem{SSBK11}
S.~Sachdev and B.~Keimer.
\newblock ``Quantum criticality",
  \href{http://dx.doi.org/10.1063/1.3554314}{\textsl{Physics Today} \textbf{64}
  29 (2011)}, \href{http://arxiv.org/abs/1102.4628}{\texttt{arXiv:1102.4628}}.

\bibitem{schmalian}
D.~E. Sheehy and J.~Schmalian.
\newblock ``Quantum critical scaling in graphene",
  \href{http://journals.aps.org/prl/abstract/10.1103/PhysRevLett.99.226803}{\textsl{Physical
  Review Letters} \textbf{99} \texttt{226803} (2007)},
  \href{http://arxiv.org/abs/0707.2945}{\texttt{arXiv:0707.2945}}.

\bibitem{muller1}
M.~M\"uller and S.~Sachdev.
\newblock ``Collective cyclotron motion of the relativistic plasma in
  graphene",
  \href{http://journals.aps.org/prb/abstract/10.1103/PhysRevB.78.115419}{\textsl{Physical
  Review} \textbf{B78} \texttt{115419} (2008)},
  \href{http://arxiv.org/abs/0801.2970}{\texttt{arXiv:0801.2970}}.

\bibitem{muller4}
L.~Fritz, J.~Schmalian, M.~M\"uller, and S.~Sachdev.
\newblock ``Quantum critical transport in clean graphene",
  \href{http://journals.aps.org/prb/abstract/10.1103/PhysRevB.78.085416}{\textsl{Physical
  Review} \textbf{B78} \texttt{085416} (2008)},
  \href{http://arxiv.org/abs/0802.4289}{\texttt{arXiv:0802.4289}}.

\bibitem{andrei}
G.~Li, A.~Lucian, and E.~Y. Andrei.
\newblock ``Scanning tunneling spectroscopy of graphene on graphite",
  \href{http://journals.aps.org/prl/abstract/10.1103/PhysRevLett.102.176804}{\textsl{Physical
  Review Letters} \textbf{102} \texttt{176804} (2009)},
  \href{http://arxiv.org/abs/0803.4016}{\texttt{arXiv:0803.4016}}.

\bibitem{lanzara}
D.~A. Siegel, W.~Regan, A.~V. Fedorov, A.~Zettl, and A.~Lanzara.
\newblock ``Charge-carrier screening in single-layer graphene",
  \href{http://journals.aps.org/prl/abstract/10.1103/PhysRevLett.110.146802}{\textsl{Physical
  Review Letters} \textbf{110} \texttt{146802} (2013)},
  \href{http://arxiv.org/abs/1304.5837}{\texttt{arXiv:1304.5837}}.

\bibitem{damle97}
K.~{Damle} and S.~{Sachdev}.
\newblock ``Nonzero-temperature transport near quantum critical points",
  \href{http://dx.doi.org/10.1103/PhysRevB.56.8714}{\textsl{Physical Review}
  \textbf{B56} 8714 (1997)},
  \href{http://arxiv.org/abs/cond-mat/9705206}{\texttt{arXiv:cond-mat/9705206}}.

\bibitem{sachdev}
S.~Sachdev.
\newblock \emph{Quantum Phase Transitions}
  \href{http://www.amazon.com/Quantum-Phase-Transitions-Subir-Sachdev/dp/0521514681/ref=sr_1_1?ie=UTF8&qid=1433031213&sr=8-1&keywords=quantum+phase+transitions}{(Cambridge
  University Press, $2^{\mathrm{nd}}$ ed., 2011)}.

\bibitem{hkms}
S.~A. Hartnoll, P.~K. Kovtun, M.~M\"uller, and S.~Sachdev.
\newblock ``Theory of the Nernst effect near quantum phase transitions in
  condensed matter, and in dyonic black holes",
  \href{http://journals.aps.org/prb/abstract/10.1103/PhysRevB.76.144502}{\textsl{Physical
  Review} \textbf{B76} \texttt{144502} (2007)},
  \href{http://arxiv.org/abs/0706.3215}{\texttt{arXiv:0706.3215}}.

\bibitem{muller2}
M.~M\"uller, L.~Fritz, and S.~Sachdev.
\newblock ``Quantum-critical relativistic magnetotransport in graphene",
  \href{http://journals.aps.org/prb/abstract/10.1103/PhysRevB.78.115406}{\textsl{Physical
  Review} \textbf{B78} \texttt{115406} (2008)},
  \href{http://arxiv.org/abs/0805.1413}{\texttt{arXiv:0805.1413}}.

\bibitem{muller3}
M.~M\"uller, J.~Schmalian, and L.~Fritz.
\newblock ``Graphene -- a nearly perfect fluid",
  \href{http://journals.aps.org/prl/abstract/10.1103/PhysRevLett.103.025301}{\textsl{Physical
  Review Letters} \textbf{103} \texttt{025301} (2009)},
  \href{http://arxiv.org/abs/0903.4178}{\texttt{arXiv:0903.4178}}.

\bibitem{foster}
M.~S. Foster and I.~L. Aleiner.
\newblock ``Slow imbalance relaxation and thermoelectric transport in
  graphene",
  \href{http://journals.aps.org/prb/abstract/10.1103/PhysRevB.79.085415}{\textsl{Physical
  Review} \textbf{B79} \texttt{085415} (2009)},
  \href{http://arxiv.org/abs/0810.4342}{\texttt{arXiv:0810.4342}}.

\bibitem{dsz}
R.~A. Davison, K.~Schalm, and J.~Zaanen.
\newblock ``Holographic duality and the resistivity of strange metals",
  \href{http://journals.aps.org/prb/abstract/10.1103/PhysRevB.89.245116}{\textsl{Physical
  Review} \textbf{B89} \texttt{245116} (2014)},
  \href{http://arxiv.org/abs/1311.2451}{\texttt{arXiv:1311.2451}}.

\bibitem{fong}
K.~C. Fong and K.~C. Schwab.
\newblock ``Ultrasensitive and wide-bandwidth thermal measurements of graphene
  at low temperatures",
  \href{http://journals.aps.org/prb/abstract/10.1103/PhysRevX.2.031006}{\textsl{Physical
  Review} \textbf{X2} \texttt{031006} (2012)},
  \href{http://arxiv.org/abs/1202.5737}{\texttt{arXiv:1202.5737}}.

\bibitem{fong2}
K.~C. Fong, E.~Wollman, H.~Ravi, W.~Chen, A.~Clerk, M.~D. Shaw, H.~D. LeDuc,
  and K.~C. Schwab.
\newblock ``Measurement of the electronic thermal conductance channels and heat
  capacity of graphene at low temperature",
  \href{http://journals.aps.org/prx/abstract/10.1103/PhysRevX.3.041008}{\textsl{Physical
  Review} \textbf{X3} \texttt{041008} (2013)},
  \href{http://arxiv.org/abs/1308.2265}{\texttt{arXiv:1308.2265}}.

\bibitem{crossno2}
J.~Crossno, X.~Liu, T.~A. Ohki, P.~Kim, and K.~C. Fong.
\newblock ``Development of high frequency and wide bandwidth Johnson noise
  thermometry",
  \href{http://scitation.aip.org/content/aip/journal/apl/106/2/10.1063/1.4905926}{\textsl{Applied
  Physics Letters} \textbf{106} \texttt{023121} (2015)},
  \href{http://arxiv.org/abs/1411.4596}{\texttt{arXiv:1411.4596}}.

\bibitem{crossno}
J.~Crossno \emph{et al.}
\newblock ``Observation of the Dirac fluid and the breakdown of the
  Wiedemann-Franz law in graphene",
  \href{http://arxiv.org/abs/1509.04713}{\texttt{arXiv:1509.04713}}.

\bibitem{lucas}
A.~Lucas.
\newblock ``Hydrodynamic transport in strongly coupled disordered quantum field
  theories", \href{http://arxiv.org/abs/1506.02662}{\texttt{arXiv:1506.02662}}.

\bibitem{btv}
M.~Blake, D.~Tong, and D.~Vegh.
\newblock ``Holographic lattices give the graviton an effective mass",
  \href{http://journals.aps.org/prl/abstract/10.1103/PhysRevLett.112.071602}{\textsl{Physical
  Review Letters} \textbf{112} \texttt{071602} (2014)},
  \href{http://arxiv.org/abs/1310.3832}{\texttt{arXiv:1310.3832}}.

\bibitem{herzog}
K.~Balasubramanian and C.~P. Herzog.
\newblock ``Losing forward momentum holographically",
  \href{http://iopscience.iop.org/0264-9381/31/12/125010}{\textsl{Classical and
  Quantum Gravity} \textbf{31} \texttt{125010} (2014)},
  \href{http://arxiv.org/abs/1312.4953}{\texttt{arXiv:1312.4953}}.

\bibitem{lss}
A.~Lucas, S.~Sachdev, and K.~Schalm.
\newblock ``Scale-invariant hyperscaling-violating holographic theories and the
  resistivity of strange metals with random-field disorder",
  \href{http://journals.aps.org/prd/abstract/10.1103/PhysRevD.89.066018}{\textsl{Physical
  Review} \textbf{D89} \texttt{066018} (2014)},
  \href{http://arxiv.org/abs/1401.7993}{\texttt{arXiv:1401.7993}}.

\bibitem{lucas1501}
A.~Lucas.
\newblock ``Conductivity of a strange metal: from holography to memory
  functions",
  \href{http://link.springer.com/article/10.1007%2FJHEP03%282015%29071}{\textsl{Journal
  of High Energy Physics} \textbf{03} \texttt{071} (\textbf{2015})},
  \href{http://arxiv.org/abs/1501.05656}{\texttt{arXiv:1501.05656}}.

\bibitem{davison15}
R.~A. Davison and B.~Gout\'eraux.
\newblock ``Dissecting holographic conductivities",
  \href{http://link.springer.com/article/10.1007%2FJHEP09%282015%29090}{\textsl{Journal
  of High Energy Physics} \textbf{09} \texttt{090} (\textbf{2015})},
  \href{http://arxiv.org/abs/1505.05092}{\texttt{arXiv:1505.05092}}.

\bibitem{blake2}
M.~Blake.
\newblock ``Momentum relaxation from the fluid/gravity correspondence",
  \href{http://link.springer.com/article/10.1007%2FJHEP09%282015%29010}{\textsl{Journal
  of High Energy Physics} \textbf{09} \texttt{010} (\textbf{2015})},
  \href{http://arxiv.org/abs/1505.06992}{\texttt{arXiv:1505.06992}}.

\bibitem{donos1506}
A.~Donos and J.~P. Gauntlett.
\newblock ``Navier-Stokes on black hole horizons and DC thermoelectric
  conductivity",
  \href{http://arxiv.org/abs/1506.01360}{\texttt{arXiv:1506.01360}}.

\bibitem{grozdanov}
S.~Grozdanov, A.~Lucas, S.~Sachdev, and K.~Schalm.
\newblock ``Absence of disorder-driven metal-insulator transitions in simple
  holographic models",
  \href{http://arxiv.org/abs/1507.00003}{\texttt{arXiv:1507.00003}}.

\bibitem{yacoby2007}
J.~Martin, N.~Akerman, G.~Ulbricht, T.~Lohmann, J.~H. Smet, K.~von Klitzing,
  and A.~Yacoby.
\newblock ``Observation of electron-hole puddles in graphene using a scanning
  single electron transistor",
  \href{http://www.nature.com/nphys/journal/v4/n2/full/nphys781.html}{\textsl{Nature
  Physics} \textbf{4} 144 (2008)},
  \href{http://arxiv.org/abs/0705.2180}{\texttt{arXiv:0705.2180}}.

\bibitem{sarmachargepuddles}
E.~Rossi and S.~Das Sarma.
\newblock ``Ground-state of graphene in the presence of random charged
  impurities",
  \href{http://journals.aps.org/prl/abstract/10.1103/PhysRevLett.101.166803}{\textsl{Physical
  Review Letters} \textbf{101} \texttt{166803} (2008)},
  \href{http://arxiv.org/abs/0803.0963}{\texttt{arXiv:0803.0963}}.

\bibitem{crommie}
Y.~Zhang, V.~W. Brar, C.~Girit, A.~Zettl, and M.~F. Crommie.
\newblock ``Origin of spatial charge inhomogeneity in graphene",
  \href{http://www.nature.com/nphys/journal/v5/n10/abs/nphys1365.html}{\textsl{Nature
  Physics} \textbf{5} 722 (2009)},
  \href{http://arxiv.org/abs/0902.4793}{\texttt{arXiv:0902.4793}}.

\bibitem{xue}
J.~Xue, J.~Sanchez-Yamagishi, D.~Bulmash, P.~Jacquod, A.~Deshpande,
  K.~Watanabe, T.~Taniguchi, P.~Jarillo-Herrero, and B.~J. LeRoy.
\newblock ``Scanning tunnelling microscopy and spectroscopy of ultra-flat
  graphene on hexagonal boron nitride",
  \href{http://www.nature.com/nmat/journal/v10/n4/full/nmat2968.html}{\textsl{Nature
  Materials} \textbf{10} 282 (2011)},
  \href{http://arxiv.org/abs/1102.2642}{\texttt{arXiv:1102.2642}}.

\bibitem{sarma1}
S.~Adam, E.~H. Hwang, V.~Galitski, and S.~Das Sarma.
\newblock ``A self-consistent theory for graphene transport",
  \href{http://www.pnas.org/content/104/47/18392}{\textsl{Proceedings of the
  National Academy of Sciences} \textbf{104} 18392 (2007)},
  \href{http://arxiv.org/abs/0705.1540}{\texttt{arXiv:0705.1540}}.

\bibitem{sarma2}
E.~Rossi, S.~Adam, and S.~Das Sarma.
\newblock ``Effective medium theory for disordered two-dimensional graphene",
  \href{http://journals.aps.org/prb/abstract/10.1103/PhysRevB.79.245423}{\textsl{Physical
  Review} \textbf{B79} \texttt{245423} (2009)},
  \href{http://arxiv.org/abs/0809.1425}{\texttt{arXiv:0809.1425}}.

\bibitem{galitski}
D.~K. Efimkin and V.~Galitski.
\newblock ``A strongly-interacting Dirac liquid on the surface of a topological
  Kondo insulator",
  \href{http://journals.aps.org/prb/abstract/10.1103/PhysRevX.2.031006}{\textsl{Physical
  Review} \textbf{B90} \texttt{081113} (2014)},
  \href{http://arxiv.org/abs/1404.5640}{\texttt{arXiv:1404.5640}}.

\bibitem{soljacic}
L.~Lu, Z.~Wang, D.~Ye, L.~Ran, L.~Fu, J.~D. Joannoupoulos, and
  M.~Solja\v{c}i\'c.
\newblock ``Experimental observation of Weyl points",
  \href{http://www.sciencemag.org/content/349/6248/622}{\textsl{Science}
  \textbf{349} 622 (2015)},
  \href{http://arxiv.org/abs/1502.03438}{\texttt{arXiv:1502.03438}}.

\bibitem{syxu}
S.-Y.~Xu \emph{et al.}
\newblock ``Discovery of a Weyl fermion semimetal and topological Fermi arcs",
  \href{http://www.sciencemag.org/content/349/6248/613}{\textsl{Science}
  \textbf{349} 613 (2015)},
  \href{http://arxiv.org/abs/1502.03807}{\texttt{arXiv:1502.03807}}.

\bibitem{bqlv}
B.~Q.~Lv \emph{et al.}
\newblock ``Experimental discovery of Weyl semimetal TaAs",
  \href{http://journals.aps.org/prx/abstract/10.1103/PhysRevX.5.031013}{\textsl{Physical
  Review} \textbf{X5} \texttt{031013} (2015)},
  \href{http://arxiv.org/abs/1502.04684}{\texttt{arXiv:1502.04684}}.

\bibitem{nielsen}
H.~N. Nielsen and M.~Ninomiya.
\newblock ``The Adler-Bell-Jackiw anomaly and Weyl fermions in a crystal",
  \href{http://www.sciencedirect.com/science/article/pii/0370269383915290}{\textsl{Physics
  Letters} \textbf{B130} 389 (1983)}.

\bibitem{spivakson}
D.~T. Son and B.~Z. Spivak.
\newblock ``Chiral anomaly and classical negative magnetoresistance of Weyl
  metals",
  \href{http://journals.aps.org/prb/abstract/10.1103/PhysRevB.88.104412}{\textsl{Physical
  Review} \textbf{B88} \texttt{104412} (2012)},
  \href{http://arxiv.org/abs/1206.1627}{\texttt{arXiv:1206.1627}}.

\bibitem{goldsmid}
G.~S. Nolas and H.~J. Goldsmid.
\newblock ``Thermal conductivity of semiconductors", in \emph{Thermal
  Conductivity: Theory, Properties and Applications} (ed. T. M. Tritt), 105,
  \href{http://www.amazon.com/Quantum-Phase-Transitions-Subir-Sachdev/dp/0521514681/ref=sr_1_1?ie=UTF8&qid=1433031213&sr=8-1&keywords=quantum+phase+transitions}{(Kluwer
  Academic, 2004)}.

\bibitem{yoshino}
H.~Yoshino and K.~Murata.
\newblock ``Significant enhancement of electronic thermal conductivity of
  two-dimensional zero-gap systems by bipolar-diffusion effect",
  \href{http://journals.jps.jp/doi/abs/10.7566/JPSJ.84.024601}{\textsl{Journal
  of the Physical Society of Japan} \textbf{84} \texttt{024601} (2015)}.

\bibitem{hartnollads}
S.~A. Hartnoll.
\newblock ``Lectures on holographic methods for condensed matter physics",
  \href{http://iopscience.iop.org/0264-9381/26/22/224002/}{\textsl{Classical
  and Quantum Gravity} \textbf{26} \texttt{224002} (2009)},
  \href{http://arxiv.org/abs/0903.3246}{\texttt{arXiv:0903.3246}}.

\bibitem{vafek}
O.~Vafek.
\newblock ``Anomalous thermodynamics of Coulomb-interacting massless Dirac
  fermions in two spatial dimensions",
  \href{http://journals.aps.org/prl/abstract/10.1103/PhysRevLett.98.216401}{\textsl{Physical
  Review Letters} \textbf{98} \texttt{216401} (2007)},
  \href{http://arxiv.org/abs/cond-mat/0701145}{\texttt{arXiv:cond-mat/0701145}}.

\bibitem{geim2005}
K.~S. Novoselov, A.~K. Geim, S.~V. Morozov, D.~Jiang, M.~I. Katsnelson, I.~V.
  Grigorieva, S.~V. Dubonos, and A.~A. Firsov.
\newblock ``Two-dimensional gas of massless Dirac fermions in graphene",
  \href{http://www.nature.com/nature/journal/v438/n7065/full/nature04233.html}{\textsl{Nature}
  \textbf{438} 197 (2005)}.

\bibitem{kim2005}
Y.~Zhang, Y.~W. Tan, H.~L. Stormer, and P.~Kim.
\newblock ``Experimental observation of the quantum Hall effect and Berry's
  phase in graphene",
  \href{http://www.nature.com/nature/journal/v438/n7065/full/nature04235.html}{\textsl{Nature}
  \textbf{438} 201 (2005)}.

\bibitem{breusing}
M.~Breusing, C.~Ropers, and T.~Elsaesser.
\newblock ``Ultrafast carrier dynamics in graphite",
  \href{http://journals.aps.org/prl/abstract/10.1103/PhysRevLett.102.086809}{\textsl{Physical
  Review Letters} \textbf{102} \texttt{086809} (2009)}.

\bibitem{johannsen}
J.~C.~Johannsen \emph{et al}.
\newblock ``Direct view of hot carrier dynamics in graphene",
  \href{http://journals.aps.org/prl/abstract/10.1103/PhysRevLett.111.027403}{\textsl{Physical
  Review Letters} \textbf{111} \texttt{027403} (2013)},
  \href{http://arxiv.org/abs/1304.2615}{\texttt{arXiv:1304.2615}}.

\bibitem{sarma2009}
E.~H. Hwang and S.~Das Sarma.
\newblock ``Screening-induced temperature-dependent transport in
  two-dimensional graphene",
  \href{http://journals.aps.org/prb/abstract/10.1103/PhysRevB.79.165404}{\textsl{Physical
  Review} \textbf{B79} \texttt{165404} (2009)},
  \href{http://arxiv.org/abs/0811.1212}{\texttt{arXiv:0811.1212}}.

\bibitem{lucasMM}
A.~Lucas and S.~Sachdev.
\newblock ``Memory matrix theory of magnetotransport in strange metals",
  \href{http://journals.aps.org/prb/abstract/10.1103/PhysRevB.91.195122}{\textsl{Physical
  Review} \textbf{B91} \texttt{195122} (2015)},
  \href{http://arxiv.org/abs/1502.04704}{\texttt{arXiv:1502.04704}}.

\bibitem{wang13}
L.~Wang \emph{et al.}
\newblock ``One-dimensional electrical contact to a two-dimensional material",
  \href{http://www.sciencemag.org/content/342/6158/614}{\textsl{Science}
  \textbf{342} 614 (2013)}.

\bibitem{kss}
P.~Kovtun, D.~T. Son, and A.~O. Starinets.
\newblock ``Viscosity in strongly interacting quantum field theories from black
  hole physics",
  \href{http://journals.aps.org/prl/abstract/10.1103/PhysRevLett.94.111601}{\textsl{Physical
  Review Letters} \textbf{94} \texttt{111601} (2005)},
  \href{http://arxiv.org/abs/hep-th/0405231}{\texttt{arXiv:hep-th/0405231}}.

\bibitem{cao}
C.~Cao, E.~Elliott, J.~Joseph, H.~Wu, J.~Petricka, T.~Sch\"afer, and J.~E.
  Thomas.
\newblock ``Universal quantum viscosity in a unitary Fermi gas",
  \href{http://www.sciencemag.org/content/331/6013/58.full}{\textsl{Science}
  \textbf{331} 58 (2011)},
  \href{http://arxiv.org/abs/1007.2625}{\texttt{arXiv:1007.2625}}.

\bibitem{luzum}
M.~Luzum and P.~Romatschke.
\newblock ``Conformal relativistic viscous hydrodynamics: applications to RHIC
  results at $\sqrt{s_{\mathrm{NN}}}=200$ GeV",
  \href{http://journals.aps.org/prc/abstract/10.1103/PhysRevC.78.034915}{\textsl{Physical
  Review} \textbf{C78} \texttt{034915} (2008)},
  \href{http://arxiv.org/abs/0804.4015}{\texttt{arXiv:0804.4015}}.

\bibitem{huard}
B.~Huard, N.~Stander, J.~A. Sulpizio, and D.~Goldhaber-Gordon.
\newblock ``Evidence of the role of contacts on the observed electron-hole
  asymmetry in graphene",
  \href{http://journals.aps.org/prb/abstract/10.1103/PhysRevB.78.121402}{\textsl{Physical
  Review} \textbf{B78} \texttt{121402} (2008)},
  \href{http://arxiv.org/abs/0804.2040}{\texttt{arXiv:0804.2040}}.

\bibitem{kirkpatrick}
S.~Kirkpatrick.
\newblock ``Classical transport in disordered media: scaling and
  effective-medium theory",
  \href{http://journals.aps.org/prl/abstract/10.1103/PhysRevLett.27.1722}{\textsl{Physical
  Review Letters} \textbf{27} 1722 (1971)}.

\bibitem{derrida}
B.~Derrida and J.~Vannimenus.
\newblock ``A transfer-matrix approach to random resistor networks",
  \href{http://iopscience.iop.org/0305-4470/15/10/007}{\textsl{Journal of
  Physics} \textbf{A15} L557 (1982)}.

\bibitem{hwang07}
E.~H. Hwang and S.~Das Sarma.
\newblock ``Acoustic phonon scattering limited carrier mobility in 2D extrinsic
  graphene",
  \href{http://journals.aps.org/prb/abstract/10.1103/PhysRevB.77.115449}{\textsl{Physical
  Review} \textbf{B77} \texttt{115449} (2008)},
  \href{http://arxiv.org/abs/0711.0754}{\texttt{arXiv:0711.0754}}.

\bibitem{efetov}
D.~K. Efetov and P.~Kim.
\newblock ``Controlling electron-phonon interactions in graphene at ultrahigh
  carrier densities",
  \href{http://journals.aps.org/prl/abstract/10.1103/PhysRevLett.105.256805}{\textsl{Physical
  Review Letters} \textbf{105} \texttt{256805} (2010)},
  \href{http://arxiv.org/abs/1009.2988}{\texttt{arXiv:1009.2988}}.

\bibitem{bolotin}
K.~I. Bolotin, K.~J. Sikes, Z.~Jiang, M.~Klima, G.~Fudenberg, J.~Hone, P.~Kim,
  and H.~L. Stormer.
\newblock ``Ultrahigh electron mobility in suspended graphene",
  \href{http://www.sciencedirect.com/science/article/pii/S0038109808001178}{\textsl{Solid
  State Communications} \textbf{146} 351 (2008)},
  \href{http://arxiv.org/abs/0802.2389}{\texttt{arXiv:0802.2389}}.

\bibitem{mayorov}
A.~S. Mayorov, D.~C. Elias, I.~S. Mukhin, S.~V. Morozov, L.~A. Ponomarenko,
  K.~S. Novoselov, A.~K. Geim, and R.~V. Gorbachev.
\newblock ``How close can one approach the Dirac point in graphene
  experimentally?",
  \href{http://pubs.acs.org/doi/abs/10.1021/nl301922d}{\textsl{Nano Letters}
  \textbf{12} 4629 (2012)},
  \href{http://arxiv.org/abs/1206.3848}{\texttt{arXiv:1206.3848}}.

\bibitem{horng}
J.~Horng \emph{et al.}
\newblock ``Drude conductivity of Dirac fermions in graphene",
  \href{http://journals.aps.org/prb/abstract/10.1103/PhysRevB.83.165113}{\textsl{Physical
  Review} \textbf{B83} \texttt{165113} (2011)},
  \href{http://arxiv.org/abs/1007.4623}{\texttt{arXiv:1007.4623}}.

\bibitem{lucas2}
A.~Lucas and S.~Sachdev.
\newblock ``Transport in inhomogeneous quantum critical fluids at finite
  frequency and magnetic field", to appear.

\bibitem{ghahari}
F.~Ghahari, H-Y. Xie, T.~Tanaguchi, K.~Watanabe, M.~S. Foster, and P.~Kim.
\newblock ``Enhanced thermoelectric power in graphene: Violation of the Mott
  relation by inelastic scattering",
  \href{http://arxiv.org/abs/1601.05859}{\texttt{arXiv:1601.05859}}.

\bibitem{trefethen}
L.~N. Trefethen.
\newblock \emph{Spectral Methods in MATLAB},
  \href{http://www.amazon.com/Spectral-Methods-MATLAB-Software-Environments/dp/0898714656/ref=sr_1_1?ie=UTF8&qid=1439317114&sr=8-1&keywords=spectral+methods+in+matlab}{(SIAM,
  2000)}.

\bibitem{domaindec}
B.~Smith, P.~Bjorstad, and W.~Gropp.
\newblock \emph{Domain Decomposition: Parallel Multilevel Methods for Elliptic
  Partial Differential Equations}
  \href{http://www.amazon.com/Domain-Decomposition-Multilevel-Differential-Equations/dp/0521602866}{(Cambridge
  University Press, 2004)}.

\end{thebibliography}
\end{document}